\documentclass[10pt,journal,compsoc]{IEEEtran}
\usepackage[normalem]{ulem}
\usepackage{cite}
\usepackage{lineno,hyperref}
\usepackage{booktabs}
\modulolinenumbers[5]
\usepackage{color}
\usepackage{xcolor}
\usepackage{comment}
\usepackage{verbatim}
\usepackage{float}
\usepackage{tikz}
\usetikzlibrary[shapes,shapes.arrows,arrows,shapes.misc,fit,positioning]
\usepackage[export]{adjustbox}
\usepackage{caption}
\usepackage{hyphenat}
\usepackage{subcaption}
\usepackage{siunitx}
\usepackage{amsmath}

\tikzstyle{vertex}=[circle,fill=black!25,minimum size=20pt,inner sep=0pt]
\tikzstyle{star vertex} = [star,star points=5, star point ratio=2.25,fill=red!50,minimum size=20pt,inner sep=0pt]%
\tikzstyle{edge} = [draw,thick,-]
\tikzstyle{weight} = [font=\small]
\tikzstyle{selected edge} = [draw,line width=2pt,-,red!70]
\tikzstyle{ignored edge} = [draw,line width=5pt,-,black!20]

\usepackage{listings}
\newif\ifargonnereport
\argonnereportfalse
\newif\iffinal
\finaltrue

\iffinal
  \newcommand\todd[1]{}
  \newcommand\todo[1]{}
\else 
  \definecolor{puce}{rgb}{0.8,0.53,0.6}
  \newcommand\todd[1]{\textcolor{puce}{Todd: #1}}
  \newcommand\todo[1]{\textcolor{red}{Todo: #1}}
\fi 

        {\begin{list}{\labelitemi}{
                \setlength{\topsep}{0pt}
                \setlength{\parskip}{0pt}
                \setlength{\itemsep}{0pt}
                \setlength{\parsep}{0pt}
                \setlength{\leftmargin}{23pt}
                \setlength{\labelwidth}{23pt}
        }}
        {\end{list}}

{\tiny }
\begin{document}

\ifargonnereport
\onecolumn
\pagestyle{empty}

\vspace{1.75in}

\begin{centering}

ARGONNE NATIONAL LABORATORY

9700 South Cass Avenue

Argonne, Illinois  60439

\vspace{1.5in}

{\large \textbf{The PetscSF Scalable Communication Layer}}

\vspace{.5in}

\textbf{Junchao Zhang, Jed Brown, Satish Balay, Jacob Faibussowitsch, Matthew Knepley, Oana Marin, Richard Tran Mills, Todd Munson, Barry F. Smith, Stefano Zampini}

\vspace{.5in}

Mathematics and Computer Science Division

\vspace{.25in}

Preprint ANL/MCS-P9449-0221

\vspace{.5in}

February 2021

\end{centering}

\vspace{2.0in}

\bigskip

\par\noindent
\footnotetext [1]
{
This work was supported by the Exascale Computing Project (17-SC-20-SC), a collaborative effort of the U.S.\ Department of Energy Office of Science and the National Nuclear Security Administration, and by the U.S.\ Department of Energy under Contract DE-AC02-06CH11357 and Office of Science Awards DE-SC0016140 and DE-AC02-0000011838.
This research used resources of the Oak Ridge Leadership Computing Facilities, a DOE Office of Science User Facility supported under Contract DE-AC05-00OR22725.
}

\newpage

\vspace*{\fill}
\begin{center}
\fbox{
\parbox{4in}{
The submitted manuscript has been created by UChicago Argonne, LLC, Operator of Argonne
National Laboratory (``Argonne''). Argonne, a U.S. Department of Energy Office of Science
laboratory, is operated under Contract No. DE-AC02-06CH11357. The U.S. Government retains
for itself, and others acting on its behalf, a paid-up nonexclusive, irrevocable worldwide
license in said article to reproduce, prepare derivative works, distribute copies to the
public, and perform publicly and display publicly, by or on behalf of the Government.
The Department of Energy will provide public access to these results of federally
sponsored research in accordance with the DOE Public Access
Plan. \url{http://energy.gov/downloads/doe-public-accessplan}
}}
\end{center}
\vfill

\newpage
\twocolumn
\pagestyle{plain}
\setcounter{page}{1}

\fi

\title{The PetscSF Scalable Communication Layer}

\author{
\IEEEauthorblockN{
  Junchao Zhang\IEEEauthorrefmark{1},		%
  Jed Brown\IEEEauthorrefmark{2},		%
  Satish Balay\IEEEauthorrefmark{1},		%
  Jacob Faibussowitsch\IEEEauthorrefmark{3},
  Matthew Knepley\IEEEauthorrefmark{4},		%
  Oana Marin\IEEEauthorrefmark{1},		%
  Richard Tran Mills\IEEEauthorrefmark{1},	%
  Todd Munson\IEEEauthorrefmark{1},		%
  Barry F. Smith\IEEEauthorrefmark{5},		%
  Stefano Zampini\IEEEauthorrefmark{6}		%
}

\IEEEauthorblockA{
\IEEEauthorrefmark{1}Argonne National Laboratory, 
Lemont, IL 60439 USA
}

\IEEEauthorblockA{
\IEEEauthorrefmark{2}University of Colorado Boulder, Boulder, CO 80302 USA
}

\IEEEauthorblockA{
\IEEEauthorrefmark{3}University of Illinois at Urbana-Champaign, Urbana, IL 61801 USA
}

\IEEEauthorblockA{
\IEEEauthorrefmark{4}University at Buffalo, Buffalo, NY 14260 USA
}

\IEEEauthorblockA{
\IEEEauthorrefmark{5}Argonne Associate of Global Empire, LLC, Argonne National Laboratory, 
Lemont, IL 60439 USA
}

\IEEEauthorblockA{
\IEEEauthorrefmark{6}King Abdullah University of Science and Technology, 
Thuwal, Saudi Arabia
}
}

\maketitle

\begin{abstract}
PetscSF, the communication 
component of the Portable, Extensible Toolkit for Scientific Computation (PETSc), 
is designed to provide PETSc's communication infrastructure suitable for exascale
computers that utilize GPUs and other accelerators.
PetscSF provides a simple application programming interface (API) for managing  common communication 
patterns in scientific computations by using a star-forest 
graph representation.
PetscSF supports several implementations based on MPI and NVSHMEM,
whose selection is
 based on the characteristics of the application or the target architecture. An efficient and portable model for network and intra-node communication is essential for implementing large-scale applications.
The Message Passing 
Interface, which has been the de facto standard for distributed memory systems,
has developed into a large complex API that does not yet provide high performance on
the emerging heterogeneous CPU-GPU-based exascale systems.
In this paper, we discuss the design of PetscSF, how it can overcome some difficulties of working directly with  MPI on GPUs, and we demonstrate its performance, scalability, and novel features.
\end{abstract}

\begin{IEEEkeywords}
Communication, GPU, extreme-scale, MPI, PETSc
\end{IEEEkeywords}

\section{Introduction} \label{sec:introduction}

\IEEEPARstart{D}{istributed} memory computation is practical and scalable; all high-end supercomputers today are distributed memory systems.
Orchestrating the communication between processes with separate memory spaces is an essential
part of programming such systems. 
Programmers need interprocess communication to coordinate the processes' work, 
distribute data, manage data dependencies, and balance 
workloads. 
The Message Passing Interface (MPI) is the de facto standard for communication on distributed 
memory systems.
Parallel applications and libraries in scientific computing are predominantly 
programmed in MPI.
However, writing communication code directly with MPI primitives, especially in applications with irregular 
communication patterns, is difficult, time-consuming, and not the primary interest of application developers.
Also, MPI has not yet fully adjusted to heterogeneous CPU-GPU-based systems,
where avoiding expensive CPU-GPU synchronizations is crucial to achieving high performance.

Higher-level communication libraries tailored
for specific families of applications can significantly reduce the programming burden from the direct use of MPI.
Requirements for such libraries include an easy-to-understand communication model,
scalability, and efficient implementations while supporting a wide range of communication scenarios.
When programming directly with MPI primitives, one faces a vast number of options, for example, MPI two-sided vs. one-sided communication, individual sends and receives, neighborhood operations, persistent or non-persistent interfaces, and more.

In \cite{towards} we discuss the plans and progress in adapting the Portable, Extensible Toolkit for Scientific Computation and Toolkit for Advanced Optimization \cite{petsc-user-ref} (PETSc) to CPU-GPU systems. 
This paper focuses specifically on the plans for managing the network and intra-node communication within PETSc. 
PETSc has traditionally used MPI calls directly in the source code. 
PetscSF is the communication component we are implementing to gradually remove these direct MPI calls. Like all PETSc components, PetscSF is designed to be used both within the PETSc libraries and also by PETSc users. It is not a drop-in replacement for MPI. Though we focus our discussion on GPUs in this paper, PetscSF also supports CPU-based systems with high performance. Most of our experimental work has been done on the OLCF IBM/NVIDIA Summit system that serves as a surrogate for the future exascale systems; however, as discussed in \cite{towards}, the PetscSF design and work is focused on the emerging exascale systems.

PetscSF can do communication on any MPI datatype 
and supports a rich set of operations. 
Recently, we consolidated VecScatter (a PETSc module for communication on vectors) and PetscSF by implementing 
the scatters using PetscSF. 
We now have a single communication component in PETSc, thus making code maintenance easier. 
We added various optimizations to PetscSF, provided multiple implementations, and, most 
importantly, added GPU support. Notably, we added support of NVIDIA NVSHMEM \cite{NVSHMEM} to provide an MPI 
alternative for communication on CUDA devices.
With this, we can avoid device synchronizations needed by MPI 
and accomplish a distributed fully asynchronous computation, which is key to unleashing the power of exascale machines where
heterogeneous architecture will be the norm. 

The paper is organized as follows. Section 2 discusses related work on 
abstracting communications on distributed memory systems. 
Section 3 introduces  PetscSF.
In Section 4, we describe several examples of usage of PetscSF within PETSc itself.
In Section 5, we detail the PetscSF implementations and optimizations. 
Section 6 gives some experimental results and shows PetscSF's performance and scalability. 
Section 7 concludes the paper with a summary and look to future work.

\section{Related Work}
Because communication is such an integral part of distributed-memory programming, many regard the communication model 
{\em as} the programming model.
We compare alternative distributed memory programming models against the predominantly used MPI.

In its original (and most commonly used) form, MPI uses a two-sided communication model, where both the sender and the 
receiver are explicitly involved in the communication. Programmers directly using MPI must  manage these relationships, 
in addition to  
allocating staging buffers, determining which data to send, and finally packing and/or unpacking data as needed.
These tasks can be difficult when information about which data is shared with which processes is not explicitly known.
Significant effort has been made to overcome this drawback with mixed results. 

The High Performance Fortran (HPF) \cite{HPF} project  allowed users to write data distribution directives 
for their 
arrays and then planned for the compilers to determine the needed communication.
HPF failed because compilers, even today, are not powerful enough to do this with indirectly indexed arrays.
Several  Partitioned Global Address Space (PGAS) languages were developed,
such as UPC~\cite{bonachea2013upc}, Co-array Fortran~\cite{CAF}, Chapel~\cite{Chapel}, and OpenSHMEM~\cite{OpenSHMEM}. 
They provide 
users an illusion of shared memory.
Users can dereference global pointers, access
global arrays, or put/get remote data without the remote side's explicit participation.
Motivated by these ideas, MPI added one-sided communication in MPI-2.0 and further enhanced it in MPI-3.0.
However, the PGAS model had limited success. Codes using such models are error-prone, since shared-memory programming easily leads to data race conditions and deadlocks. These are difficult to detect, debug, and fix.
Without great care, PGAS applications are prone to low performance since programmers can easily write fine-grained 
communication code, which severely hurts performance on distributed memory systems.
Because of this, writing a correct and efficient code that requires communication using PGAS languages is not fundamentally easier than programming in MPI.

While MPI has excellent support for writing libraries and many MPI-based libraries target specific domains,
 few communication libraries are built using MPI. We surmise the reason for this is that the data to be communicated
is usually embedded in the user's data structure, and there is no agreed-upon interface to describe 
the user's data connectivity graph.
Zoltan \cite{ZoltanIsorropiaOverview2012} is one of the few such libraries that is built on MPI. 
Zoltan is a collection of data management 
services for parallel, unstructured, 
adaptive, and dynamic applications. 
Its unstructured communication service and data migration tools are close to those of PetscSF.
Users need to provide pack and unpack callback functions to Zoltan.
Not only does PetscSF pack and unpack automatically, it is capable of performing reductions when unpacking.
PetscSF supports composing communication graphs, see Section 3.3.
The gs library\cite{gs} used by Nek5000 \cite{nek5000} gathers and scatters vector entries.
It is similar to
PETSc's VecScatter, but
it is slightly more general and supports communicating multiple data types.
DIY \cite{DIY} is a block-parallel communication library for implementing scalable algorithms
that can execute both in-core and out-of-core. 
Using a block parallelism abstraction, it combines distributed-memory message passing with shared-memory thread 
parallelism. 
Complex communication patterns can be built on a graph of blocks,
usually coarse-grained, whereas in PetscSF, graph vertices can be as small as a single floating-point number.
We are unaware that any of these libraries support GPUs.

Since distributed-memory communication systems are often conflated with programming models, we should clarify that tools such as Kokkos \cite{KOKKOS} and Raja \cite{RAJA} provide functionality orthogonal to communication systems. 
For example, in \cite{towards}, we provide a prototype code that uses PetscSF for the communication and Kokkos as the portable programming model. SYCL \cite{SYCL} automatically manages needed communication between GPU and CPU but does not have general communication abilities. Thus, it also requires a distributed memory communication system.

\section{The PetscSF Interface} \label{sec:sfapi}
PetscSF has a simple conceptual model with a small, but powerful, interface. We begin by introducing how to create PetscSF objects and their primary use cases.

\subsection{PetscSF Creation} \label{sec:sfcreate}
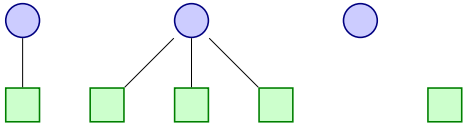
\begin{figure}
  \centering
   \resizebox{.7\linewidth}{!}{
  \begin{tikzpicture}
    \tikzstyle{every node}=[rectangle,draw=blue!50!black,fill=blue!20,thick,minimum size=6mm,rounded corners=3mm]
    \tikzstyle{every child node}=[rectangle,draw=green!50!black,fill=green!20,thick,rounded corners=0mm]
    \node at (0,0) {}
    child { node {} };
    \node at (3,0) {}
    child {node {}}
    child {node {}}
    child {node {}};
    \node at (6,0) {};
    \node[rectangle,rounded corners=0mm,draw=green!50!black,fill=green!20] at (7.5,-1.5) {};
  \end{tikzpicture}
  }
  \caption{Examples of stars, the union of which forms a star forest.
    Root vertices are identified with circles and leaves with rectangles.
    Note that roots with no leaves are allowed as well as leaves with no root. 
  }\label{fig:stars}
\end{figure}

PetscSF supports both structured and irregular communication graphs, with the latter being the focus of this paper.
A star forest is used to describe the communication graphs.
Stars are simple trees consisting of one root vertex connected to zero or more leaves. The number of leaves is 
called the \textit{degree} of the root.
We also allow leaves without connected roots. These isolated leaves represent holes in the user's
data structure that do not participate in the communication.
Fig. \ref{fig:stars} shows some example stars. A union of disjoint stars is called a star forest (SF). PetscSFs are typically partitioned across multiple processes.
In graph terminology, an PetscSF object can be regarded as being similar to a quotient 
graph \cite{saad1995data}, which embeds the closely connected subpartitions of a larger graph, which is the abstraction 
of a domain topological representation.

Following PETSc's object creation convention, 
one creates a PetscSF with \texttt{PetscSFCreate(MPI\_Comm comm,PetscSF *sf)}, where \texttt{comm} specifies the MPI 
communicator the PetscSF
lives on. One then describes the graph by calling \texttt{ 
PetscSFSetGraph()} on 
each process.
\begin{lstlisting}
typedef struct {PetscInt rank,offset;} PetscSFNode;

PetscErrorCode PetscSFSetGraph(PetscSF sf, PetscInt nroots,
  PetscInt nleaves, const PetscInt *local,
  PetscCopyMode localmode, const PetscSFNode *remote,
  PetscCopyMode remotemode);
\end{lstlisting}
This call indicates that this process owns \texttt{nroots} roots and  \texttt{nleaves} \textit{connected} leaves.
The roots are numbered from 0 to \texttt{nroots-1}.
\texttt{local[0..nleaves]} contains local indices of those connected leaves.
If \texttt{NULL} is passed for \texttt{local}, the leaf space is contiguous.
Addresses of the roots for each leaf are specified by \texttt{remote[0..nleaves]}. Leaves can connect to local or remote roots; therefore we use tuples 
\texttt{(rank,offset)} 
to represent root addresses, where \texttt{rank} is the MPI rank of the owner process of a root and \texttt{offset} is 
the index of the root on that process. 
\begin{figure}[htbp]
  \centering
  \resizebox{.7\linewidth}{!}{
  \begin{tikzpicture}[scale=0.6]
    \node[blue!50!black] (rtitle) at (0,1) {\bf Roots};
    \begin{scope}[every node/.style={rectangle,draw=blue!50!black,fill=blue!20,thick,minimum size=5mm,rounded 
    corners=2.5mm}]
      \node (r11) at (0,-1) {0}; 
      \node (r12) at (0,-2) {1}; 
      \node (r13) at (0,-3) {2}; 
      \node (r21) at (0,-4.5) {0};
      \node (r22) at (0,-5.5) {1};
      \node (r23) at (0,-6.5) {2};
      \node (r24) at (0,-7.5) {3};
      \node (r31) at (0,-9) {0};
      \node (r32) at (0,-10) {1};
    \end{scope}
    \begin{scope}[every node/.style={rectangle,draw=blue!50!black,thick}]
      \node[rectangle,fill=none,fit=(r11)(r13)] (root0) {};
      \node[rectangle,fill=none,fit=(r21)(r24)] (root1) {};
      \node[rectangle,fill=none,fit=(r31)(r32)] (root2) {};
    \end{scope}
    \node[thick,left] at (root0.west) {rank 0};
    \node[thick,left] at (root1.west) {rank 1};
    \node[thick,left] at (root2.west) {rank 2};
    \node[green!50!black] (rtitle) at (6,1) {\bf Leaves};
    \begin{scope}[every node/.style={rectangle,draw=green!50!black,fill=green!20,thick}]
    \node (c1100) at (6,0) {0}; 
    \node (c1200) at (6,-1) {1}; 
    \node (c1300) at (6,-2) {2}; 
    \node (c1400) at (6,-3) {3}; 
    \node[fill=none,fit=(c1100) (c1400)] (leaf0) {};
    \node (c2100) at (6,-4.5) {0};
    \node (c2200) at (6,-5.5) {1};
    \node (c2300) at (6,-6.5) {2};
    \node (c2400) at (6,-7.5) {3};
    \node[fill=none,fit=(c2100) (c2400)] (leaf1) {};
    \node (c3100) at (6,-9) {0};
    \node (c3200) at (6,-10) {1};
    \node (c3300) at (6,-11) {2};
    \node[fill=none,fit=(c3100) (c3300)] (leaf2) {};
    \end{scope}
    \begin{scope}[thick,draw=black!50,>=stealth]
      \draw (r11) -- (c2200);
      \draw (r11) -- (c3100);
      \draw (r13) -- (c1400);
      \draw (r21) -- (c1200);
      \draw (r21) -- (c1300);
      \draw (r21) -- (c3200);
      \draw (r23) -- (c1100);
      \draw (r24) -- (c3300);
      \draw (r31) -- (c2100);
      \draw (r32) -- (c2400);
    \end{scope}
    \node[black!70] (rtitle) at (9,1) {\tt local};
    \node[black!70] (rtitle) at (11,1) {\tt remote};
    \begin{scope}[every node/.style={rectangle,draw=black!50,fill=black!20,thick}]
    \node (ll1100) at (9,0) {0};
    \node (ll1200) at (9,-1) {1};
    \node (ll1300) at (9,-2) {2};
    \node (ll1400) at (9,-3) {3};
    \node[fill=none,fit=(ll1100) (ll1400)] (leaf0) {};
    \node (lr1100) at (11,0) {(1,2)}; 
    \node (lr1200) at (11,-1) {(1,0)}; 
    \node (lr1300) at (11,-2) {(1,0)}; 
    \node (lr1400) at (11,-3) {(0,2)}; 
    \node[fill=none,fit=(lr1100) (lr1400)] (leaf0) {};

    \node (ll2100) at (9,-4.5) {0};
    \node (ll2200) at (9,-5.5) {1};
    \node (ll2400) at (9,-6.5) {3};
    \node[fill=none,fit=(ll2100) (ll2400)] (leaf1) {};
    \node (lr2100) at (11,-4.5) {(2,0)};
    \node (lr2200) at (11,-5.5) {(0,0)};
    \node (lr2400) at (11,-6.5) {(2,1)};
    \node[fill=none,fit=(lr2100) (lr2400)] (leaf1) {};

    \node (ll3100) at (9,-9) {0};
    \node (ll3200) at (9,-10) {1};
    \node (ll3300) at (9,-11) {2};
    \node[fill=none,fit=(ll3100) (ll3300)] (leaf2) {};
    \node (lr3100) at (11,-9) {(0,0)};
    \node (lr3200) at (11,-10) {(1,0)};
    \node (lr3300) at (11,-11) {(1,3)};
    \node[fill=none,fit=(lr3100) (lr3300)] (leaf2) {};
    \end{scope}
  \end{tikzpicture}
  }
  \caption{Distributed star forest partitioned across three processes, with the specification arrays at right. Roots (leaves) on each rank are numbered top-down, with local indices starting from zero. Note the isolated vertices on rank 1. 
  }
\label{fig:starforest}
\end{figure}
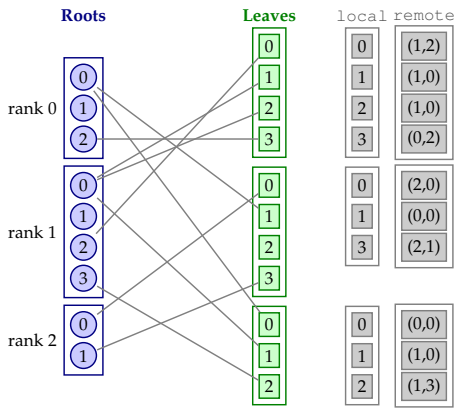
Fig. \ref{fig:starforest} shows an PetscSF with data denoted at vertices, together with the
\texttt{local} and \texttt{remote} arguments passed to \texttt{PetscSFSetGraph()}.
The edges of an PetscSF are specified by the process that owns the leaves;
therefore a given root vertex is merely a candidate for \textit{incoming} edges. This one-sided specification is
important to encode graphs containing roots with a very high degree, such as globally coupled constraints,
in a scalable way.
\texttt{PetscSFSetUp()} sets up internal data structures for its 
implementations and checks for all possible optimizations. 
The setup cost is usually amortized by multiple operations performed on the PetscSF.
A PetscSF provides only a template for communication.
Once a PetscSF is created, one can instantiate simultaneous communication
on it with different data.
All PetscSF routines return an error code; we will omit it in their prototypes
for brevity.

\subsection{PetscSF Communication Operations} \label{sec:basic_sf_ops}
In the following, we demonstrate operations that update data on vertices. All operations 
are split 
into matching \emph{begin} and \emph{end} phases, taking the same arguments. Users can insert independent computations between the calls to overlap computation and communication. Data buffers must not be altered between the two phases, and the content of the result 
buffer can be accessed only after the matching End operation has been posted. 

\vspace{5pt}
\noindent
\textbf{(1) Broadcast roots to leaves or reduce leaves to roots}

\begin{lstlisting}
PetscSFBcastBegin/End(PetscSF sf, MPI_Datatype unit, 
  const void *rootdata, void *leafdata, MPI_Op op);
PetscSFReduceBegin/End(PetscSF sf, MPI_Datatype unit, 
  const void *leafdata, void *rootdata, MPI_Op op);
\end{lstlisting}
These operations are the most used PetscSF operations (see concrete examples in Section \ref{sec:sf_matrix_uses}).
\texttt{unit} is the data type of the tree vertices. Any MPI data type can be used.
The pointers \texttt{rootdata} and \texttt{leafdata} reference 
the user's data structure organized as arrays of \texttt{unit}s. 
The term \texttt{op} designates an MPI reduction, 
such as MPI\_SUM, which tells PetcSF to add root values to their leaves' value (\texttt{SFBcast}) or 
vice versa (\texttt{SFReduce}).
When \texttt{op} is MPI\_REPLACE, the operations overwrite destination with source values.

\vspace{5pt}
\noindent
\textbf{(2) Gather leaves at roots and scatter back}\\
One may want to gather (in contrast to reduce) leaf values to roots.
For that purpose, one needs a new SF by splitting roots in the old SF into new roots as many as
their degree and then connecting leaves to the new roots one by one.
To build that SF, a difficulty is for leaves to know their new roots' offset at the remote side.
We provide a fetch-and-add operation to help this task.
\begin{lstlisting}
PetscSFFetchAndOpBegin/End(PetscSF sf,MPI_Datatype unit,
  void *rootdata,const void *leafdata,void *leafupdate,
  MPI_Op op);
\end{lstlisting}

With that, one can first do fetch-and-add on integers (\texttt{unit} = MPI\_INT) with the old SF, with roots 
initialized to their old offset value 
and leaves to 1. The operation internally does these:
For a root has $n$ leaves, $leaf_{0..n-1}$, whose values are added (\texttt{op} = MPI\_SUM) to the root in 
$n$ reductions. When updating using $leaf_i$, it first fetches the current, partially 
reduced value of the root before adding $leaf_i$'s value. 
The fetched value, which is exactly the offset of the new root $leaf_i$ is going to be connected to in the new SF, is 
stored at $leaf_i$'s position in array \texttt{leafupdate}.
After that, constructing the new SF is straightforward. 
This set of operations are so useful that we give the new SF a name, \textit{multi-SF},
and provide more user-friendly routines:
\vspace{5pt}
\noindent
\begin{lstlisting}
PetscSFGetMultiSF(PetscSF sf, PetscSF *multi);
PetscSFGatherBegin/End(PetscSF sf,MPI_Datatype unit,
  const void *leafdata,void *multirootdata);
PetscSFScatterBegin/End(PetscSF sf,MPI_Datatype unit,
  const void *multirootdata,void *leafdata);
\end{lstlisting}

The first routine returns the multi-SF of \texttt{sf} while the other two 
make use of the internal multi-SF representation of \texttt{sf}. \texttt{SFGather} gathers 
leaf values from \texttt{leafdata} and stores them at \texttt{multirootdata}, which is an array containing $m$ 
\texttt{unit}s, where  $m$ is the number of roots of the
multi-sf. \texttt{SFScatter} reverses the operation of \texttt{SFGather}.
A typical use of \texttt{SFGather}/\texttt{Scatter} in distributed computing is for owner points, working as an 
arbitrator, to 
make some non-trivial decision based on data gathered from ghost points and then scatter the decision back.

\subsection{PetscSF Composition Operations} \label{sec:sfgather}
One can make new PetscSFs from existing ones, using\\
\vspace{5pt}
\noindent \textbf{(1) Concatenation:} Suppose there are two PetscSFs, \texttt{A} and \texttt{B}, and \texttt{A}'s 
leaves are 
 overlapped with \texttt{B}'s 
roots. One call
\begin{lstlisting}
PetscSFCompose(PetscSF A,PetscSF B,PetscSF *AB);
\end{lstlisting}
 to compose a new PetscSF \texttt{AB},  whose roots are \texttt{A}'s roots and leaves are
\texttt{B}'s leaves. A root and a leaf of \texttt{AB} are connected if there is a leaf in \texttt{A} and 
a root in \texttt{B}
that bridge them. Similarly, if \texttt{A}'s leaves are overlapped with \texttt{B}'s leaves and \texttt{B}'s 
roots all have a 
degree at most one, then the result of the following composition is also well defined. 
\begin{lstlisting}
PetscSFComposeInverse(PetscSF A,PetscSF B,PetscSF* AB);
\end{lstlisting}
\texttt{AB} is a new PetscSF with  \texttt{A}'s roots and leaves being \texttt{B}'s roots and edges 
built upon  a reachability definition similar to that in  PetscSFCompose.
Such SF concatenations enable users to redistribute or re-order data from existing distributions or 
orderings.

\vspace{5pt}
\noindent
\textbf{(2) Embedding:} PetscSF allows removal of existing vertices and their associated edges, a 
common use case in scientific computations, for example, 
in field segregation in multiphysics application and subgraph extraction.
The API
\begin{lstlisting}
PetscSFCreateEmbeddedRootSF(PetscSF sf, PetscInt n,
  const PetscInt selected_roots[], PetscSF *esf);
PetscSFCreateEmbeddedLeafSF(PetscSF sf, PetscInt n,
  const PetscInt selected_leaves[], PetscSF *esf)
\end{lstlisting}
removes edges from all but the selected roots/leaves without remapping indices and returns a new PetscSF that can be used to 
perform the subcommunication using the original root/leaf data buffers.

\section{Use Cases}
Although  PetscSF has many uses,
we describe here only a small subset of applications of PetscSF to other PETSc operations, namely,  parallel matrix and 
unstructured mesh operations.

\subsection{Parallel Matrix Operations} \label{sec:sf_matrix_uses}

\begin{figure}[htbp]
\begin{center}
\includegraphics[width=.7\linewidth]{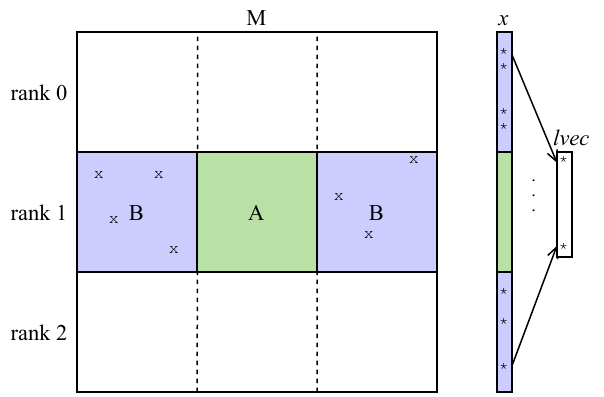}
\caption{A PETSc matrix \texttt{M} on three ranks. On rank 1 the diagonal block \texttt{A} is in green and 
the 
off-diagonal
block \texttt{B} is in blue. The x in \texttt{B} denotes nonzeros, while the * in \textit{x} denotes remote vector entries needed 
to  
compute $Bx$.}
\label{fig:mpiaij}
\end{center}
\end{figure}

{\bf Sparse matrix-vector products (SpMV)}  \label{sec:matmult} The standard parallel sparse matrix implementation in PETSc distributes a matrix by rows.
On each MPI rank, a block of consecutive rows makes up a local matrix.
The parallel vectors that can be multiplied by the matrix are distributed by rows in a conforming manner.
On each rank, PETSc \textit{splits} the
local matrix
into two blocks: a \textit{diagonal} block $A$, whose columns 
match with the 
vector's rows on the current rank, and an \textit{off-diagonal} block $B$ for the remaining columns. See Fig. 
\ref{fig:mpiaij}.
$A$ and $B$ are separately encoded in the  compressed sparse row (CSR) format with \textit{reduced local} column 
indices.
Then, the SpMV in PETSc is decomposed into $y = Ax + Bx$. $Ax$  
depends solely on 
local vector data, 
while $Bx$ requires access to remote entries of $x$ that correspond to the nonzero columns of 
$B$. 
PETSc uses a 
sequential vector \textit{lvec} to hold these needed ghost points entries. The length of \textit{lvec} is equal to the 
number of nonzero columns of $B$,
and its entries are ordered in their corresponding column order. See Fig. \ref{fig:mpiaij}.

We build an PetscSF for the communication needed in $Bx$ as described below.
On each rank, there are $n$ roots, where $n$ is the local size of  \textit{x}, and 
$m$ leaves, where $m$ is the size of vector \textit{lvec}.
The leaves have contiguous indices running from 0 to $m-1$, each connected to a root representing a global column index.
With the matrix layout info available on each rank, we  calculate the owner rank and local index of a global column on its owner and determine the PetscSFNode argument of SFSetGraph() introduced in Section \ref{sec:sfcreate}.
The SpMV $y=Mx$ can be implemented in the listing below,
which overlaps the local computation $y=Ax$ with the communication.

\begin{lstlisting}
// x_d, lvec_d are data arrays of x, lvec respectively
PetscSFBcastBegin(sf,MPI_DOUBLE,x_d,lvec_d,MPI_REPLACE);
y  = A*x;
PetscSFBcastEnd(sf,MPI_DOUBLE,x_d,lvec_d,MPI_REPLACE);
y += B*lvec;
\end{lstlisting}

The transpose multiply $y=M^Tx$ is implemented by
\begin{lstlisting}
y    = A^T*x;
lvec = B^T*x;
PetscSFReduceBegin(sf,MPI_DOUBLE,lvec_d,y_d,MPI_SUM);
PetscSFReduceEnd(sf,MPI_DOUBLE,lvec_d,y_d,MPI_SUM);
\end{lstlisting}
Note that $lvec=B^Tx$ computes the local contributions to some remote entries of $y$. These must be added back to their
owner rank.

{\bf Extracting a submatrix from a sparse matrix} The routine \texttt{MatCreateSubMatrix(M,ir,ic,c,P)}
extracts a parallel submatrix \texttt{P} on the same set of ranks as the original matrix \texttt{M}.
Parallel global index sets, \texttt{ir} and \texttt{ic}, provide the rows the local rank should obtain and columns that 
should be in its diagonal block of the submatrix on this rank, respectively.
The difficulty is to determine, for each rank, which columns are needed (by any ranks) from the owned block of rows.

Given two SFs, \texttt{sfA}, which maps reduced local column indices (leaves) to global columns (roots) of the 
original 
matrix \texttt{M}, and \texttt{sfB}, which maps ``owned'' columns (leaves) of the submatrix \texttt{P} to global 
columns 
(roots) of  \texttt{M}, 
we determine the retained local columns  using the following:
\begin{enumerate}
\item SFReduce using \texttt{sfB}, results in a distributed array containing 
the columns  (in the form of global indices) of \texttt{P} that columns of \texttt{M} will be replicated into. 
Unneeded columns are tagged by a negative index.
\item SFBcast using \texttt{sfA} to get these values in the reduced local column space.
\end{enumerate}
The algorithm prepares the requested rows of \texttt{M} in \textit{non-split} form and distributes them to their new 
owners, which can be done using a row-based PetscSF.

\subsection{Unstructured Meshes}
PetscSF was originally introduced in \texttt{DMPlex}, the PETSc class that manages
unstructured meshes. PetscSF is one of its central data structures and is heavily used there. 
Interested readers are referred to \cite{KnepleyKarpeev09,LangeMitchellKnepleyGorman2015,hapla2021fully}.
Here we only provide a glimpse of how we use DMPlex and PetscSF together to
\textit{mechanically} build complex, distributed data distributions
and communication from simple combination operations.

DMPlex provides unstructured mesh management  
using a topological abstraction known as a Hasse diagram~\cite{HasseDiagram}, a directed acyclic graph (DAG) 
representation 
of a partially ordered set. The abstract mesh representation consists of \textit{points}, corresponding to vertices, 
edges, faces, and cells, which are connected by \textit{arrows} indicating the topological adjacency relation. 
Points from different co-dimensions are indexed uniformly, 
therefore this model is  \textit{dimension-independent};  
it can represent meshes with different cell types, including cells of 
different dimensions.

Parallel topology is defined by a mesh point PetscSF, which connects ghost points 
(leaves) to owned points (roots). 
Additionally, PetscSFs are used to represent the communication patterns needed for 
common mesh operations, such as partitioning and redistribution, ghost exchange, and assembly of discretized functions 
and operators. To generate these PetscSFs, we use \texttt{PetscSection} objects, which 
maps points to sizes of data of interest related to the points, assuming the data is packed.
For example, with an initial mesh point PetscSF, applying a 
PetscSection mapping mesh points to degrees-of-freedom (dofs) on points generates a new dof-PetscSF that relates dofs 
on 
different 
processes. Another example, with the dof-PetscSF just generated, applying  
a PetscSection mapping dofs to adjacent dofs produces 
the PetscSF for the Jacobian, described in detail in~\cite{KnepleyLangeGorman2017}.

{\bf Mesh distribution}, in response to a partition, is handled by creating several PetscSF objects representing the 
steps needed to perform the distribution. Based on the partition, we make a PetscSF whose roots are the original mesh 
points and whose leaves are the redistributed mesh points so that SFBcast would migrate the points. Next, a 
PetscSection describing topology and the original PetscSF are used to build a PetscSF to distribute the topological 
relation. Then, the PetscSections describing the data 
layout for coordinates and other mesh fields are used to build PetscSFs that redistribute data over the mesh.

{\bf Ghost communication and assembly} is also managed by PetscSFs. 
PETSc routines
\texttt{DMGlobalToLocal()} and \texttt{DMLocalToGlobal()} use the aforementioned dof-PetscSF to communicate data 
between a global vector and a local vector with ghost points.
Moreover, this communication pattern is reused to perform assembly as part of finite element and finite volume discretizations. The \texttt{PetscFE} and \texttt{PetscFV} objects can be used to 
automatically create the PetscSection objects for these data layouts, which together with the mesh point PetscSF 
automatically create the assembly PetscSF. This style of programming, with the emphasis on the declarative 
specification of data 
layout and communication patterns, \textit{automatically} generating the actual communication and data manipulation 
routines, 
has resulted in more robust, extensible, and performant code.

\section{Implementations} \label{sec:implementations}
PetscSF has multiple implementations, including ones that utilize MPI one-sided or two-sided communication.
We focus on the default implementation, which uses persistent MPI sends and receives for two-sided communication. In addition, we emphasize the implementations for GPUs. 

\subsection{Building Two-Sided Information} \label{sec:twosided}
From the input arguments of PetscSFSetGraph, each MPI rank knows which ranks (i.e., root ranks) have 
roots that the 
current rank's leaves are connected to. 
In PetscSFSetUp, we compute the reverse information: which ranks (i.e., leaf ranks) have leaves of the current rank's roots.
A typical algorithm uses MPI\_Allreduce on an integer array of the MPI communicator size, 
which is robust but non-scalable. A more scalable algorithm uses MPI\_Ibarrier, \cite{hoefler2010scalable}, which is 
PETSc's default with large communicators.
In the end, on each rank,
we compile the following information: 
\begin{enumerate}
\item A list of root ranks, connected by leaves on this rank;
\item For each root rank, a list of leaf indices representing 
leaves on 
this rank that connects to the root rank;
\item A list of leaf ranks, connected by roots on this rank;
\item For each leaf rank, a 
list of root indices, representing roots on this rank that connect to the leaf rank.
\end{enumerate}
These data structures facilitate message coalescing, which is crucial for performance on distributed memory.
Note that processes usually play double roles: they can be both a root rank and a leaf rank.

\subsection{Reducing Packing Overhead} \label{sec:packingOverhead}

With the two-sided information, we could have a simple implementation,
using \texttt{PetscSFBcast(sf, MPI\_DOUBLE, rootdata, leafdata, op)} as an example (Section 
\ref{sec:basic_sf_ops}).
Each rank, as a sender, allocates a root buffer, \texttt{rootbuf}, packs the needed root data entries 
according to the 
root indices 
(\texttt{rootidx})
obtained above into the buffer, in a form such as \texttt{rootbuf[i] = rootdata[rootidx[i]]}, and then sends data in 
\texttt{rootbuf} to leaf ranks.
Similarly, the receiver rank  allocates a leaf buffer \texttt{leafbuf} as the receive buffer. Once it has
received 
data, it 
unpacks data from 
\texttt{leafbuf} and deposits into the destination leafdata entries according to the leaf indices (\texttt{leafidx}) obtained 
above, in a form such as \texttt{leafdata[leafidx[i]] $\oplus$= leafbuf[i]}. 
Here $\oplus$= represents \texttt{op}.

However, PetscSF has several optimizations to lower or eliminate the packing (unpacking) overhead.
First, we separate local (i.e., self-to-self) and remote communications. If on a process the PetscSF has local 
edges, 
then the process  will show up in its leaf and root rank lists.
We rearrange the lists to move self to the head if that is the case. By skipping MPI for local communication, we save intermediate send and receive buffers and pack and unpack calls. 
Local communication takes the form \texttt{leafdata[leafidx[i]] $\oplus$= rootdata[rootidx[i]]}.
We call this a \textit{scatter} operation.
This optimization is important in mesh repartitioning since most cells tend to stay on their current owner and 
hence local communication volume is large.

Second, we analyze the vertex indices and discover patterns that can be used to construct pack/unpack routines with fewer indirections.
The most straightforward pattern is just contiguous indices.
In that case, we can 
 use the user-supplied rootdata/leafdata as the MPI buffers without any packing or unpacking. An important 
application 
of this 
optimization is in SpMV and its transpose introduced in Section 
\ref{sec:matmult}, where 
the leaves,  the
entries in \texttt{lvec}, are contiguous. Thus, \texttt{lvec}'s data array can be directly used as the MPI receive 
buffers in
\texttt{PetscSFBcast} of PETSc's SpMV or as MPI send buffers in \texttt{PetscSFReduce} of its transpose product.
Note that, in general, we also need to consider the \texttt{MPI\_Op} involved. If it is not an assignment, then 
we must 
allocate a receive buffer before \textit{adding} it to the final destination.
Note that allocated buffers are reused for repeated PetscSF operations.

Another pattern is represented by multi-strided indices, inspired by halo exchange in stencil computation on regular domains.
In that case, ghost points living on faces of (sub)domains are either locally strided or contiguous.
Suppose we have a three-dimensional
domain of size\texttt{ [X,Y,Z]}
with points sequentially numbered in the \textit{x, y, z} order. Also, suppose that within the domain there is a subdomain of size \texttt{[dx,dy,dz]} with the index of the
first point being \texttt{start}. Then, indices of points in the subdomain can be enumerated with expression 
\texttt{start+X*Y*k+X*j+i}, for
\texttt{(i,j,k)} in \texttt{(0$\leq$i<dx,0$\leq$j<dy,0$\leq$k<dz)}.
With this utility, faces or even the interior parts of a regular domain are all such qualified subdomains.
Carrying several such parameters is enough for us to know all indices of a subdomain and for more efficient packs and unpacks on GPUs.

\subsection{GPU-Aware MPI Support}
Accelerator computation, represented by NVIDIA CUDA GPUs, brings new challenges to MPI.
Users want to communicate data on the device while MPI runs on the host,
but the MPI specification does not have a concept of device data or host data. 
In this paper's remainder, we use CUDA as an example, but the concept applies to other GPUs.
With a non-CUDA-aware MPI implementation, programmers have to copy data back and forth between device and host to do computation on the device and communication on the host.
This is a burden for programmers.
CUDA-aware MPI can ease this problem, allowing programmers to directly pass device buffers to MPI
with the same API.
This is convenient, but there is still an MPI/CUDA semantic mismatch~\cite{SNIR}.
CUDA kernels are executed asynchronously on CUDA streams, but MPI has no concept of streams.
Hence, MPI has no way to queue its operations to streams while maintaining correct
data dependence.
In practice,  before sending data, users must synchronize the device to ensure that the data in the send buffer
is ready for MPI to access at the moment of calling MPI send routines (e.g., MPI\_Send).
After receiving (e.g., MPI\_Recv), MPI synchronizes the device again to ensure that the data in the receive buffer on the GPU is ready for users to access on \textit{any} stream. 
These excessive synchronizations can impair pipelining of kernel launches. We address this issue later in the paper. 

Since rootdata/leafdata is on the device, pack/unpack routines also have  to be implemented as kernels, with associated 
vertex indices moved to the device.
The packing optimizations discussed in the preceding paragraph are more useful on GPUs because we could either remove these kernels or save device memory otherwise allocated to store indices with patterns. 
Since we use CUDA threads to unpack data from the receive buffer in \textit{parallel}, we
distinguish the case of having duplicate indices, which may lead to data races.
This is the case, for example, in the PetscSFReduce for \texttt{MatMultTranspose} (see Section 
\ref{sec:matmult}):
A single entry of the result vector \texttt{y} will likely receive contributions from multiple leaf ranks.
In this case, we use CUDA atomics, whereas we use regular CUDA instructions when no duplicated indices are present.

PetscSF APIs introduced in Section \ref{sec:sfapi} do not have stream or memory type arguments.
Internally we call cudaPointerGetAttributes() to distinguish memory spaces.
However, since this operation is expensive (around 0.3 \si{\us} per call from our experience), 
we extended PetscSF APIs to ones such as the following:
\begin{lstlisting}
PetscSFBcastWithMemTypeBegin/End(PetscSF sf, MPI_Datatype
  unit, PetscMemType rootmtype, const void *rootdata,
  PetscMemType leafmtype, void *leafdata, MPI_Op op);
\end{lstlisting}
The extra \texttt{PetscMemType} arguments tell PetscSF where the data is. PETSc vectors have such built-in info,
so that PETSc vector scatters internally use these extended APIs.

We could further extend the APIs to include stream arguments. However, since stream arguments, like C/C++ constantness, are so intrusive,
we might have to extend many APIs to pass around the stream.
We thus take another approach. PETSc has a default stream named \textit{PetscDefaultCudaStream}.
 Almost all PETSc kernels work on this stream. 
 When PETSc calls libraries such as cuBLAS and cuSPARSE, it sets their work stream to this default one.
Although PetscSF assumes that data is on the default stream, it does provide options for users to indicate that data is on other streams so that PETSc will take stricter synchronizations.
\begin{figure}[htbp]
\begin{center}
\includegraphics[width=0.5\linewidth]{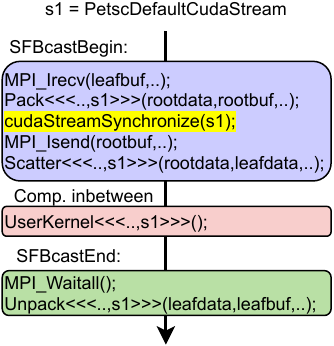}
\caption{CUDA-aware MPI support in PetscSFBcast. \textit{Scatter} denotes the local communication.
Note the cudaStreamSynchronize() before MPI\_Isend.
}
\label{fig:cudaMPI}
\end{center}
\end{figure}

Fig. \ref{fig:cudaMPI} shows a diagram of the CUDA-aware MPI support in PetscSF using PetscSFBcast as an example.
The code sequence is similar to what we have for host communication
except that pack, unpack, and scatter operations are CUDA kernels, and there is a  stream synchronization before 
MPI\_Isend, for reasons discussed above.
If input data is not on PETSc's default stream, we call
cudaDeviceSynchronize() in the beginning phase to sync
the whole device and cudaStreamSynchronize() in the end phase to sync the  
default stream so that output data is ready to be accessed afterward.

\subsection{Stream-Aware NVSHMEM Support}
\begin{figure}[htbp]
\begin{center}
\includegraphics[width=1.0\linewidth]{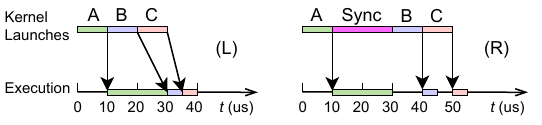}
\caption{
Pipelined kernel launches (L) vs. interrupted kernel launches (R).
 }
\label{fig:KernelLaunches}
\end{center}
\end{figure}

CUDA kernel launches have a cost of around 10 \si{\us}, which is not negligible considering that many kernels in practice have a smaller  
execution time.
An important optimization with GPU asynchronous computation is to overlap kernel launches with kernel executions 
on the
GPU, so that the launch cost is effectively hidden.
However, the mandatory CUDA synchronization brought by MPI could jeopardize this optimization since it blocks 
the otherwise
nonblocking kernel launches on the host. %
See Fig. \ref{fig:KernelLaunches} for an example. Suppose a kernel 
  launch takes 10 \si{\us} and 
  there are three kernels A, B, and C that take 20, 5, and 5 \si{\us} to execute, respectively. 
  If the kernels are launched in a nonblocking way (Fig. \ref{fig:KernelLaunches}(L)), the 
 total 
 cost
 to run them is 40 \si{\us}. Launch costs of B and C are completely hidden by the execution of A.
 However, if there is a synchronization after A (Fig. \ref{fig:KernelLaunches}(R)), the total cost will be 55 
 \si{\us}. Scientific codes usually have many MPI calls, implying that their kernel launches will be frequently 
 blocked by CUDA  synchronizations. 
 While the MPI community is exploring adding stream support in the MPI standard, we recently tried NVSHMEM for a remedy.
 
NVSHMEM \cite{NVSHMEM} is NVIDIA's implementation of the OpenSHMEM \cite{OpenSHMEM} specification on CUDA devices.
It supports point-to-point and collective communications between GPUs within a node or
over networks.
Communication can be initiated either on the host or on the device.
Host side APIs take a stream argument. 
NVSHMEM is a PGAS library
that provides one-sided \textit{put}/\textit{get} APIs for processes to access remote data.
While using get is possible, we focus on put-based communication.
In NVSHMEM, a process is called a processing element (PE). A set of PEs is a \textit{team}. The team containing all PEs 
is called \texttt{NVSHMEM\_TEAM\_WORLD}. PEs can query their rank in a team and the team's size.
One PE can use only one GPU.
These concepts are analogous to ranks, communicators, and \texttt{MPI\_COMM\_WORLD} in MPI. 
NVSHMEM can be used with MPI. It is natural to map one MPI rank to one PE. 
We use PEs and MPI ranks interchangeably.
With this approach, we are poised to bypass MPI to do 
communication on GPUs while keeping the rest of the PETSc code unchanged.

For a PE to access remote data, NVSHMEM uses the \textit{symmetric data object} concept.
At NVSHMEM initialization, all PEs allocate a block of CUDA memory, which is 
called a symmetric heap. Afterwards, every remotely accessible object  has to be \textit{collectively} 
allocated/freed by \textit{all} PEs in \texttt{NVSHMEM\_TEAM\_WORLD}, with the \textit{same} size, 
so that such an object always appears symmetrically on all PEs, at the same offset in their symmetric heap.
PEs access remote data by referencing a symmetric address and the rank of the remote PE. 
A symmetric address is the address of a symmetric object on the local PE, plus an offset if needed.
The code below allocates two symmetric double arrays \texttt{src[1]} and \texttt{dst[2]},
and every PE puts a double from its \texttt{src[0]} to the next PE's \texttt{dst[1]}.
\begin{lstlisting}[language=c,numbers=none]
double *src = nvshmem_malloc(sizeof(double));
double *dst = nvshmem_malloc(sizeof(double)*2);
int    pe   = nvshmem_team_my_pe(team); // get my rank
int    size = nvshmem_team_n_pes(team), next = (pe+1)%
nvshmemx_double_put_on_stream(&dst[1],src,1,next,stream);
\end{lstlisting}
For PEs to know the arrival of data put by other PEs and then read it, they can call a collective 
nvshmem\_barrier(team) to separate 
the put and the read,
or senders can send signals to receivers for checking. Signals are merely symmetric objects of type uint64\_t. 
We prefer the latter approach since collectives are unfit for sparse neighborhood communications that are important to PETSc.
Because of the collective memory allocation constraint, we support NVSHMEM only on PetscSFs built on 
\texttt{PETSC\_COMM\_WORLD}, which is the MPI communicator we used to initialize PETSc and NVSHMEM.

\begin{figure}[htbp]
\begin{center}
\includegraphics[width=1.0\linewidth]{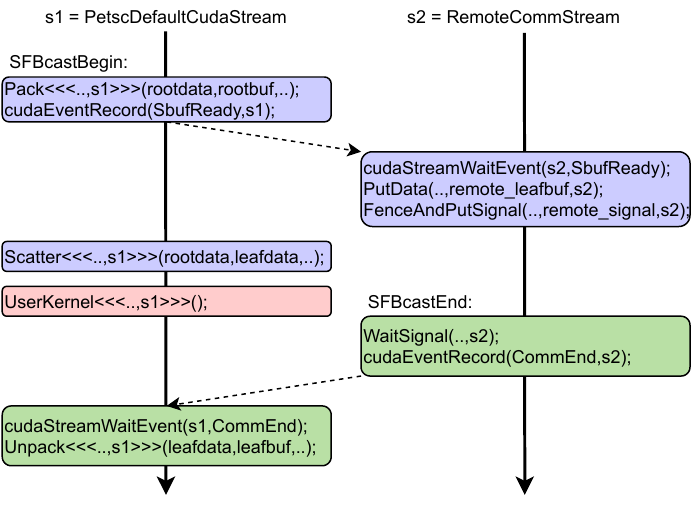}
\caption{
Stream-aware NVSHMEM support in PetscSFBcast. Blue boxes are in the beginning phase, and green boxes are in the end phase.
The red box in between is the user code. Dashed lines represent data dependence between streams.
Functions are ordered vertically and called asynchronously.
 }
\label{fig:SFNvshmem}
\end{center}
\end{figure}

Fig. \ref{fig:SFNvshmem} gives a skeleton of the NVSHMEM support in PetscSF, again using PetscSFBcast as an example.
We create a new stream \texttt{RemoteCommStream} (s2) to take charge of remote communication, 
such that communication and user's computation, denoted by \texttt{UserKernel}, could be overlapped.
First, on \texttt{PetscDefaultCudaStream} (s1), we record a CUDA event \texttt{SbufReady} right after the pack kernel
to indicate data in the send buffer is ready for send.
Before sending, stream s2 waits for the event so that the send-after-pack dependence is enforced.
Then PEs put data and \textit{end-of-put} signals to destination PEs.
To ensure that signals are delivered \textit{after} data, we do an NVSHMEM memory fence at the local PE before putting signals.
In the end phase, PEs wait until end-of-put signals targeting them have arrived (e.g., through 
\texttt{nvshmem\_uint64\_wait\_until\_all()}).
Then they record an event \texttt{CommEnd} indicating end of communication on s2.
 PEs just need to wait for that event on s1 before unpacking data from the receive buffer. 
Note that all function calls in Fig. \ref{fig:SFNvshmem} are asynchronous.

NVSHMEM provides users a mechanism to distinguish locally accessible and remotely accessible PEs.
One can roughly think of the former as intranode PEs and the latter as internode PEs.
We take advantage of this and use different NVSHMEM APIs when putting data.
We call the host API \texttt{nvshmemx\_putmem\_nbi\_on\_stream()} for each local PE,
and the device API \texttt{nvshmem\_putmem\_nbi()} on CUDA threads, with each targeting a remote PE.
For local PEs, the host API uses the efficient CUDA device copy engines to do GPUDirect peer-to-peer memory 
copy,
while for remote PEs, it uses slower GPUDirect RDMA.

\begin{figure}[htbp]
\begin{center}
\includegraphics[width=0.7\linewidth]{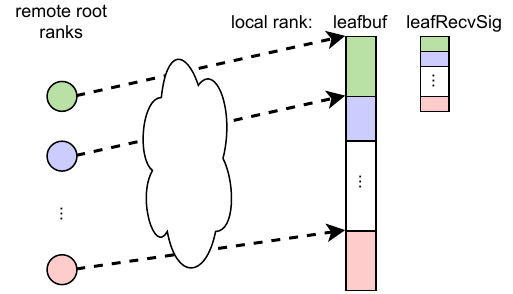}
\caption{
Right: two symmetric objects on a local rank. Left:  remote root ranks putting data to the objects. 
Root ranks need to know offsets at the remote side.
The cloud indicates these are remote accesses. 
 }
\label{fig:onesided}
\end{center}
\end{figure}

We now detail how we set up send and receive buffers for NVSHMEM.
In CUDA-aware MPI support of PetscSF, we generally need to allocate on each rank two CUDA buffers, \texttt{rootbuf} 
and 
\texttt{leafbuf}, to function as MPI 
send or receive buffers. 
Now we must allocate them symmetrically to make them accessible to NVSHMEM.
To that end,
we call an MPI\_Allreduce to get their maximal size over the communicator and  use the result in 
\texttt{nvshmem\_malloc()}.
As usual, \texttt{leafbuf} is logically split into chunks of various sizes (see Fig. \ref{fig:onesided}).
Each chunk in an PetscSFBcast operation is used as a receive buffer for an associated root rank.
Besides \texttt{leafbuf}, we allocate a symmetric object \texttt{leafRecvSig[]},
which is an array of the \textit{end-of-put} signals with each 
entry 
associated 
with a root rank.
In one-sided programming, a root rank has to know the associated chunk's offset in \texttt{leafbuf} to put the data
and also the associated signal's offset in \texttt{leafRecvSig[]} to set the signal. The preceding  explanations apply to 
\texttt{rootbuf} 
and leaf ranks similarly.
In PetscSFSetUp, we use MPI two-sided to assemble the information needed for NVSHMEM one-sided.
At the end of the setup, on each rank, we have the following new data structures in addition to those introduced in 
Section \ref{sec:twosided}:

\begin{enumerate}
\item A list of offsets, each associated with a leaf rank, showing where the 
local rank should 
send (put) its \texttt{rootbuf} data to these leaf ranks' \texttt{leafbuf}
\item A list of offsets, each associated with a leaf rank, showing where the
 local rank should set signals on these leaf ranks 

\item A list of offsets, each associated with a root rank, showing where the 
local rank should 
send (put) its \texttt{leafbuf} data to these root ranks' \texttt{rootbuf}
\item A list of offsets, each associated with a root rank, showing where the
 local rank should set signals on these root ranks
\end{enumerate}

\begin{figure}[htbp]
\begin{center}
\includegraphics[width=1.0\linewidth]{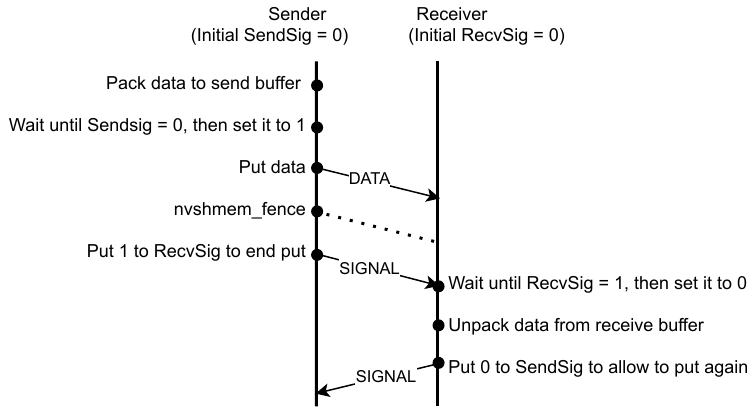}
\caption{Protocol of the PetscSF NVSHMEM put-based communication. 
The dotted line indicates that the signal below is observable only after the remote put is completed at the destination.
We have rootSendSig, rootRecvSig, leafSendSig, and leafRecvSig symmetric objects on each PE, with an initial value 0.
 }
\label{fig:protocal}
\end{center}
\end{figure}

With these, we are almost ready to implement PetscSF with  NVSHMEM.
But there is a new complexity coming with one-sided communication.
Suppose we do communication in a loop.
When receivers are still using their receive buffer, senders could move into the next iteration and put new data into the receivers' buffer and corrupt it.
To avoid this situation, we designed a protocol shown in Fig. \ref{fig:protocal}, between a pair of PEs (sender and receiver).
Besides the end-of-put signal (RecvSig) on the receiver side, 
we allocate an \textit{ok-to-put} signal (SendSig) on the sender side.
The sender must wait until the variable is 0 to begin putting in the data.
Once the receiver has unpacked data from its receive buffer, it sends 0 to the sender's SendSig
to give it  permission to put the next collection of data.

\section{Experimental Results}

We evaluated PetscSF on the Oak Ridge Leadership Computing Facility (OLCF) Summit supercomputer as a surrogate for the upcoming exascale computers.
Each  node of Summit has two sockets, each with an IBM Power9 CPU accompanied by three  
NVIDIA  Volta  V100  GPUs.
Each CPU and its three GPUs are connected by the NVLink interconnect at a 
bandwidth of 50  GB/s.  Communication  between  the  two  
CPUs  is  provided  by IBM's X-Bus at  a  bandwidth of 64 GB/s.  Each CPU also connects through a PCIe Gen4 x8 bus with a  
Mellanox InfiniBand network interface card (NIC) with a 
bandwidth of 16 GB/s.  The NIC has an injection bandwidth of 25 GB/s.
On the software side, we used gcc 6.4 and the CUDA-aware IBM Spectrum MPI 10.3.
We also used 
NVIDIA CUDA Toolkit 10.2.89, NVSHMEM 2.0.3, NCCL 2.7.8 (NVIDIA Collective Communication 
Library, which implements multi-GPU and multi-node collectives for NVIDIA GPUs),
and GDRCopy 2.0 (a low-latency GPU memory copy library). NVSHMEM needs the latter two.

In \cite{sf-tech-report}, we provided a series of microbenchmarks for PetscSF
and studied various part of it with CUDA-aware MPI. In this paper, we focus more on application-level
evaluation on both CPUs and GPUs.

\subsection{PetscSF Ping Pong Test}
To determine the overhead of PetscSF, 
we wrote a ping pong test \textit{sf\_pingpong} in PetscSF,
to compare PetscSF performance against raw MPI performance.
The test uses two MPI ranks and a PetscSF with $n$ roots on rank 0 and $n$ leaves on rank 1. 
The leaves are connected to the roots consecutively.
With this PetscSF and op = \texttt{MPI\_REPLACE}, PetscSFBcast sends a message from rank 0 to rank 1, and a following 
SFReduce  
bounces a message back.
Varying $n,$ we can then measure the latency of various message sizes,
mimicing the \textit{osu\_latency} test from the OSU microbenchmarks \cite{OSUMicro}.
By comparing the performance attained by the two tests, we can determine the overhead of PetscSF.

We measured PetscSF MPI latency
with either host-to-host messages (H-H) or device-to-device  (D-D) messages.
By device messages, we mean regular CUDA memory data, not NVSHMEM symmetric memory data.
Table \ref{tab:pingpong} shows the intra-socket results, where the two MPI ranks
were bound to the same CPU and used two GPUs associated with that CPU.
The roots and leaves in this PetscSF are contiguous,
 so PetscSF's optimization skips the pack/unpack operations. 
Thus this table compares a raw MPI code with a PetscSF code that embodies much richer semantics.
Comparing the H-H latency, we see that PetscSF has an overhead from 0.6 to 1.0 \si{\us},
 which is spent on 
 checking input arguments and bookkeeping.
The D-D latency is interesting.
It shows that PetscSF has an overhead of about 5 \si{\us} over the OSU test. 
We verified this was because PetscSF calls cudaStreamSynchronize() before sending data, whereas the OSU test does not.
We must have the synchronization in actual application codes, as discussed before.
We also performed inter-socket and inter-node tests. 
Results, see \cite{sf-tech-report}, indicated a similar gap between the PetscSF test and the OSU test.

\begin{table}[htbp]
\caption{Intra-socket host-to-host (H-H) latency and device-to-device (D-D) latency
 (\si{\us}) measured by osu\_latency (OSU) and  sf\_pingpong (SF), 
 with IBM Spectrum MPI.
}
\label{tab:pingpong}
\begin{tabular}{|r|r|r|r|r|r|r|r|}
\hline
Msg (Bytes) & 1K   & 4K   & 16K  & 64K  & 256K & 1M   & 4M    \\ \hline
OSU H-H         & 0.8  & 1.3  & 3.5  & 4.7  & 12.2 & 36.3 & 152.4 \\ \hline
SF H-H          & 1.5  & 1.9  & 4.2  & 5.5  & 12.9 & 37.3 & 151.8 \\ \hline
OSU D-D         & 17.7 & 17.7 & 17.8 & 18.4 & 22.5 & 39.2 & 110.3 \\ \hline
SF D-D          & 22.8 & 23.0 & 22.9 & 23.5 & 27.7 & 46.3 & 111.8 \\ \hline
\end{tabular}
\end{table}

\begin{figure}[htbp]
\begin{center}
\includegraphics[width=1.0\linewidth]{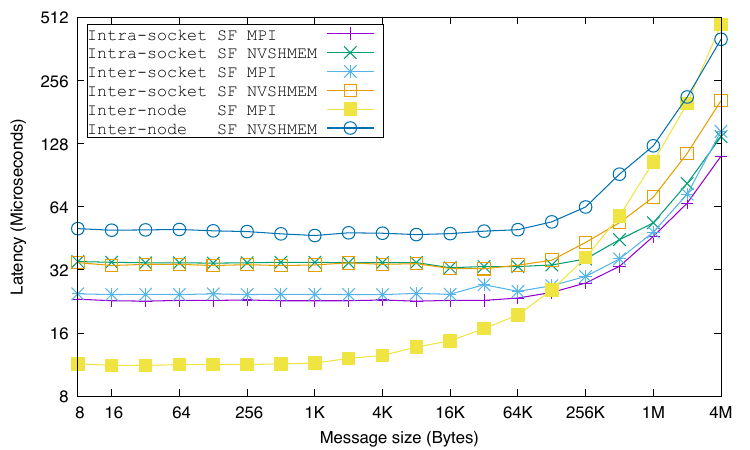}
\caption{Device to device (D-D) latency measured by
sf\_pingpong, using CUDA-aware IBM MPI or NVSHMEM.
 }
\label{fig:SFGPU}
\end{center}
\end{figure}

We used the same test and compared PetscSF MPI and PetscSF NVSHMEM, 
with results shown in Fig. \ref{fig:SFGPU}.
For small messages, NVSHMEM's latency is  about 12 \si{\us} higher than MPI's in intra-socket and 
inter-socket cases and 40 \si{\us} more in the inter-node cases.
For large inter-socket messages, the gap between the two is even larger (up to 60 \si{\us} at 4 MB).
Possible reasons for the performance difference include:
(1) In PetscSF NVSHMEM, there are extra memory copies of data between the root/leafdata in CUDA memory and the root and 
leaf buffers in NVSHMEM memory;
(2) The current NVSHMEM API has limitations. For example, we would like to use
fewer kernels to implement the protocol in Fig. \ref{fig:protocal}. Within a node,
the NVSHMEM host API delivers much better performance than its device API, forcing us to do signal-wait through 
device API, but data-put through the host API.
For another example, NVSHMEM provides a device API to do data-put and signal-put in one call, but there is no
host counterpart. One has to use two kernel launches for this task using the host API for data-put.
All these extra kernel launches increase the latency;
(3) NVSHMEM is still a new NVIDIA product. There is much headroom for it to grow.

\subsection{Asynchronous Conjugate Gradient on GPUs}
To explore distributed asynchronous execution on GPUs enabled by NVSHMEM, we adapted 
CG, the Krylov conjugate gradient solver in PETSc, to a prototype asynchronous version  CGAsync.
Key differences between the two are as follows. (1) A handful of PETSc routines they call are different. There are two categories.
The first includes routines with scalar output parameters, for example, vector dot product.
 CG calls \texttt{VecDot(Vec x,Vec y,double *a)} with \texttt{a} pointing to a host buffer, 
while CGAsync calls \texttt{VecDotAsync(Vec x, Vec y, double *a)} with \texttt{a} referencing
a device buffer.
In VecDot, each process calls cuBLAS routines
to compute a partial dot product and then copies it back to the host, where it calls
MPI\_Allreduce to get the final dot product and stores it at the host buffer.
Thus VecDot  synchronizes the  CPU and the GPU device. While in VecDotAsync,
once the partial dot product from cuBLAS is computed, each process  calls an NVSHMEM reduction operation on PETSc's
default stream to compute
the final result and stores it at the device buffer.
The second category of differences includes routines with scalar input parameters, such as \texttt{VecAXPY(Vec y,double a, Vec x)}, which 
computes 
y += a*x. CG calls \texttt{VecAXPY} while CGAsync calls \texttt{VecAXPYAsync(Vec y,double *a, Vec x)} with 
\texttt{a} 
referencing device memory, so that \texttt{VecAXPYAsync} can be queued to a  stream while \texttt{a} is computed on the device. (2) CG does scalar arithmetic (e.g., divide two scalars) on the CPU, while CGAsync does them with 
tiny \textit{scalar kernels} on the GPU.
(3) CG checks convergence (by comparison) in every iteration on the CPU to determine whether it should exit the loop while 
CGAsync does not. 
Users need 
to specify maximal iterations. This could be improved by checking for convergence every few (e.g., 20) iterations. We leave this as future work.

\begin{figure}[htbp]
\begin{center}
\subfloat{\includegraphics[width=1.0\linewidth]{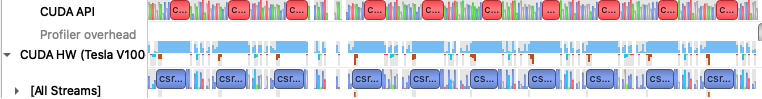}}\\
\vspace{10pt}
\subfloat{\includegraphics[width=1.0\linewidth]{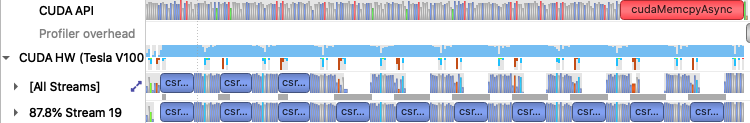}}
\caption{Timeline of CG (top) and CGAsync (bottom) on rank 2. Each ran ten iterations.
The blue \textit{csr...} bars are csrMV (i.e., SpMV) kernels in cuSPARSE, and the red \textit{c...} bars are 
cudaMemcpyAsync() copying data from device to host.
}
\label{fig:CG}
\end{center}
\end{figure}

We tested CG and CGAsync without preconditioning on a single Summit compute node with two sparse matrices from
 the SuiteSparse Matrix Collection \cite{Florida}.
The CG was run with PetscSF CUDA-aware MPI, and CGAsync was run with PetscSF NVSHMEM.
The first matrix is Bump\_2911 with about 3M rows and 128M nonzero entries.
We ran both algorithms 10 iterations with 6 MPI ranks and one GPU per rank.
Fig. \ref{fig:CG} shows their timeline through the profiler NVIDIA NSight Systems.
The kernel launches (label \texttt{CUDA API}) in CG were spread over the 
10 iterations. The reason is that in each iteration, there are multiple MPI calls 
(mainly in MatMult as discussed in Section \ref{sec:matmult},
and vector dot and norm operations),
which constantly block the kernel launch
pipeline. In CGAsync, however, while the device was executing the 8th iteration (with profiling),
the host had launched \textit{all} kernels for 10 iterations.  The long red bar cudaMemcpyAsync indicates that after the kernel launches,  the host was idle, waiting for the final result from the device.

Test results show that the time per iteration for CG and CGAsync was about 690 and 676 \si{\us},
respectively. CGAsync gave merely a 2\% improvement. 
This small improvement is because the matrix is huge, and computation is using the vast majority of the time.
From profiling, we knew SpMV alone (excluding communication)
took 420 \si{\us}. If one removes the computational time, the improvement in communication time is substantial.  Unfortunately, because of bugs in the NVSHMEM library with multiple nodes, 
we could not scale the test to more compute nodes.
Instead, we used a smaller matrix, Kuu, of about 7K rows and 340K nonzero entries to see how
CGAsync would perform in a strong-scaling sense. We repeated the above tests. 
Time per iteration for CG and CGAsync was about 300 and 250 \si{\us}.
CGAsync gave a 16.7\% improvement. 
Note that this improvement was attained despite the higher ping pong latency of PetscSF NVSHMEM.

We believe CGAsync has considerable potential for improvement.
As the NVSHMEM library matures, it should reach or surpass MPI's ping pong performance.  
We also note that there are many kernels in CGAsync; for the scalar kernels mentioned above,   kernel launch times
could not be entirely hidden with small matrices. 
We are investigating techniques like CUDA Graphs to automatically fuse kernels in the loop to further
reduce the launch cost; with MPI, such fusion within an iteration is not possible due to the synchronizations that MPI mandates.

\subsection{Mesh Distribution on CPU}

This section reports on the robustness and efficiency of the PetscSF infrastructure as used in the mesh distribution 
process through DMPlex.
In the first stage, the cell-face connectivity graph is constructed and partitioned, followed by 
the mesh data's actual migration (topology, labels associated with mesh points, and cell coordinates) and then the 
distributed mesh's final setup.
We do not analyze the stage around graph partitioning and instead focus on the timings associated with the distribution of the needed mesh data followed by the final local setup.

\begin{figure}[htbp]
\begin{center}
\includegraphics[width=0.8\linewidth]{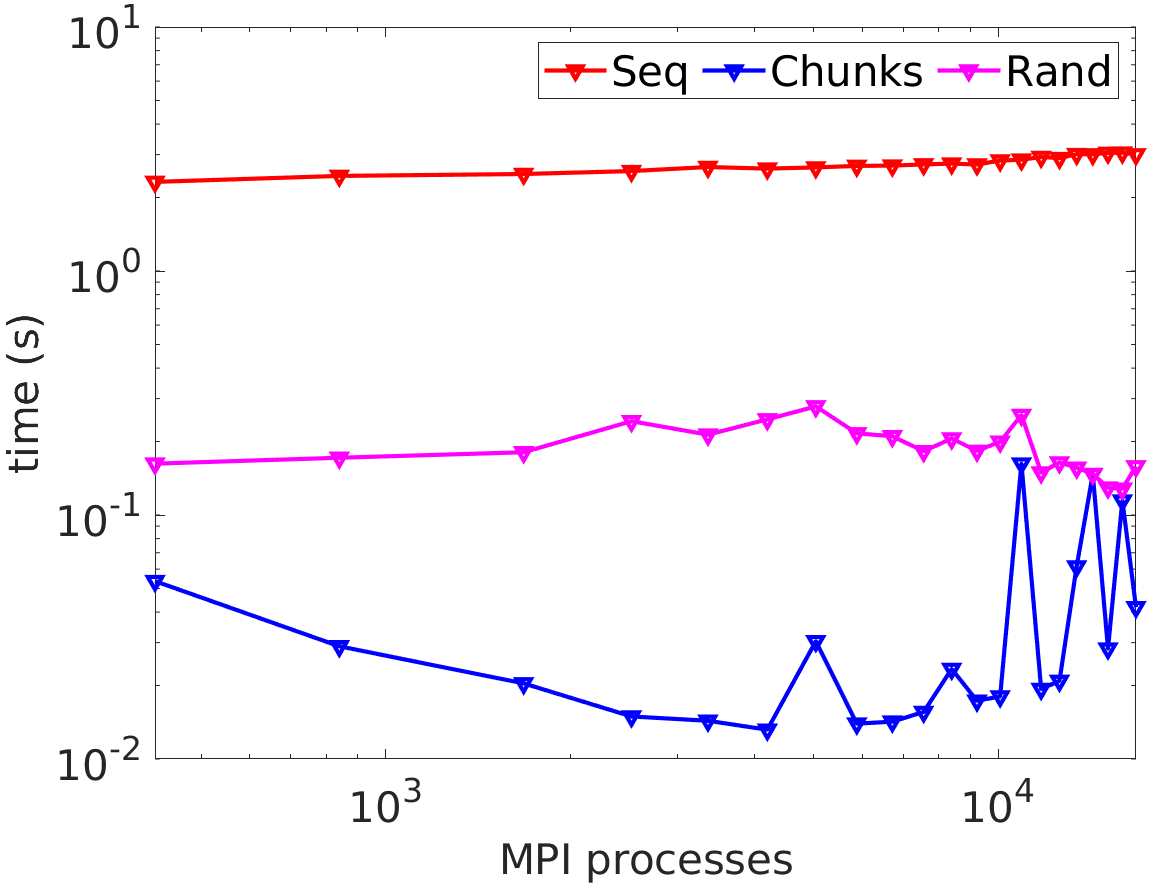}
\caption{DMPlex mesh migration timings as a function of the number of MPI processes for the Sequential, Chunks, and 
Random use cases,
showing the robustness and scalability of the mesh operations using PetscSF.
}
\label{fig:SFPlexMig}
\end{center}
\end{figure}

We consider the migration induced by a graph partitioning algorithm on
three different initial distributions of a fully periodic $128\times 128 \times 128$ hexahedral mesh:
\begin{itemize}
    \item {Seq}: the mesh is entirely stored on one process.
    \item {Chunks}: the mesh is stored in non-overlapping chunks obtained by a simple distribution of the lexicographically ordered cells.
    \item {Rand}: the mesh is stored randomly among processes.
\end{itemize}
The sequential case is common in scientific applications when the mesh is stored in a format that is not suitable for parallel reading. It features a one-to-all communication pattern.  The Chunks and Rand cases represent different mesh distribution scenarios after parallel read, and a many-to-many communication pattern characterizes them. 
Ideally, the Chunks case would have a more favorable communication pattern than the Rand case, where potentially all processes need to send/receive data from all processes. Fig. \ref{fig:SFPlexMig} collects the mesh migration timing as a 
function of the number of processes used in the distribution. Timings remain essentially constant as the number of 
processes is increased from 420 to 16,800 due to increased communication times and a decrease in the subsequent local 
setup time, confirming the scalability of the overall implementation. 
The Chunks' timings had some irregularity, since initial mesh distributions in this case varied from 
less-optimal to almost-optimal, affecting the timed second redistribution.

\subsection{Parallel Sparse Matrix-Matrix Multiplication (SpMM)}

We recently developed a generic driver for parallel sparse matrix-matrix multiplications that takes advantage of the block-row distribution used in PETSc for their storage; see Fig. \ref{fig:mpiaij}. Here we report preliminary results using cuSparse to store and manipulate the local matrices. In particular, given two parallel sparse matrices $A$ and $P$, we consider
the matrix product $A P$ and the projection operation $P^TAP$, by splitting them into three different phases:
\begin{enumerate}
    \item Collect rows of $P$ corresponding to the nonzero columns of the off-diagonal portion of $A$.
    \item Perform local matrix-matrix operations.
    \item Assemble the final product, possibly involving off-process values insertion.
\end{enumerate}
Step 1 is a submatrix extraction operation, while step 2 corresponds to a sequence of purely local matrix-matrix products that can execute on the device. We set up an intermediate PetscSF with leaf vertices represented by the row indices of the result matrix and root vertices given by its row distribution during the symbolic phase. The communication of the off-process values needed in step 3 is then performed using the \texttt{SFGather} operation on a temporary GPU buffer, followed by local GPU-resident assembly.

\begin{figure}[htbp]
\begin{center}
\includegraphics[width=1.0\linewidth]{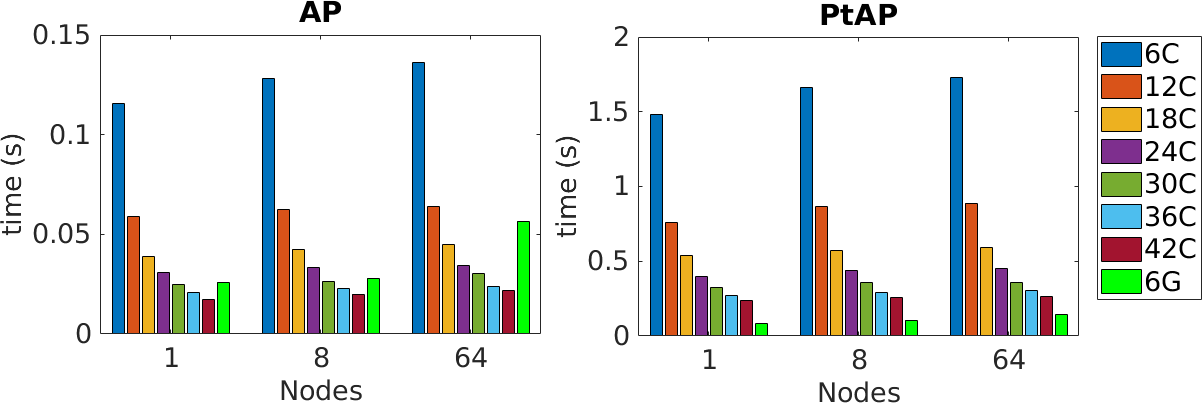}
\caption{Timings for parallel sparse matrix-matrix numerical products with constant workload per node. Left is $AP$; 
right is $P^T A P$. 6 GPUs per 
node (6G) are compared against an increasing number of cores per node (6C-42C) with different numbers of Summit nodes.}
\label{fig:SFMatMatGPU}
\end{center}
\end{figure}

The performances of the numerical phase of these two matrix-matrix operations on GPUs are plotted in Fig. 
\ref{fig:SFMatMatGPU} for  different numbers of nodes of Summit, and they are compared against our highly optimized CPU 
matrix-matrix operations directly calling MPI send and receive routines; the $A$ operators are obtained from a second-order finite element approximation of the 
Laplacian, while the $P$ matrices are generated by the native 
algebraic multigrid solver in PETSc. The finite element problem defined on a tetrahedral unstructured mesh is 
partitioned and weakly scaled among Summit nodes, from 1 to 64; the number of rows of $A$ ranges from 1.3 million to 89.3 million.
While the workload is kept fixed per node, a strong-scaling analysis is carried out within a node, and the timings 
needed by the numerical 
algorithm using 6 GPUs per node (label 6G) are compared against an increasing number of cores per node (from 6 to 42, 
labeled 6C and 42C, respectively). The Galerkin triple matrix product shows the most promising speedup, while the performances of 
the matrix product $AP$, cheaper to be computed, are more dependent on the number of nodes.
We plan to perform further analysis and comparisons 
when our NVSHMEM backend for PetscSF supports multinode configurations and we have full support for 
asynchronous device operations using streams within the PETSc library. 

\section{Conclusion and Future Work}
We introduced PetscSF, the communication component in PETSc,
including its programming model, its API, and its
implementations. We emphasized the implementation on GPUs since one of our primary goals is to provide highly efficient PetscSF implementations for the upcoming exascale computers.
Our experiments demonstrated PetscSF's performance, overhead, scalabilty,
and novel asynchronous features. 
We plan to continue to optimize PetscSF for exascale systems and to investigate asynchronous computation on GPUs 
enabled by PetscSF at large scale.

\section*{Acknowledgments}
\addcontentsline{toc}{section}{Acknowledgments}
\noindent
We thank Akhil Langer and Jim Dinan from the NVIDIA NVSHMEM team for their
assistance.
This work was supported by the Exascale Computing Project (17-SC-20-SC), a collaborative effort of the U.S.\ Department of Energy Office of Science and the National Nuclear Security Administration, and by the U.S.\ Department of Energy under Contract DE-AC02-06CH11357 and Office of Science Awards DE-SC0016140 and DE-AC02-0000011838.
This research used resources of the Oak Ridge Leadership Computing Facilities, a DOE Office of Science User Facility supported under Contract DE-AC05-00OR22725.

\bibliographystyle{IEEEtran}
\bibliography{IEEEabrv,petsc-ecp-tpds}

\vspace{-50pt}
\begin{IEEEbiography}[{\includegraphics[width=1in,height=1.25in,clip,keepaspectratio]{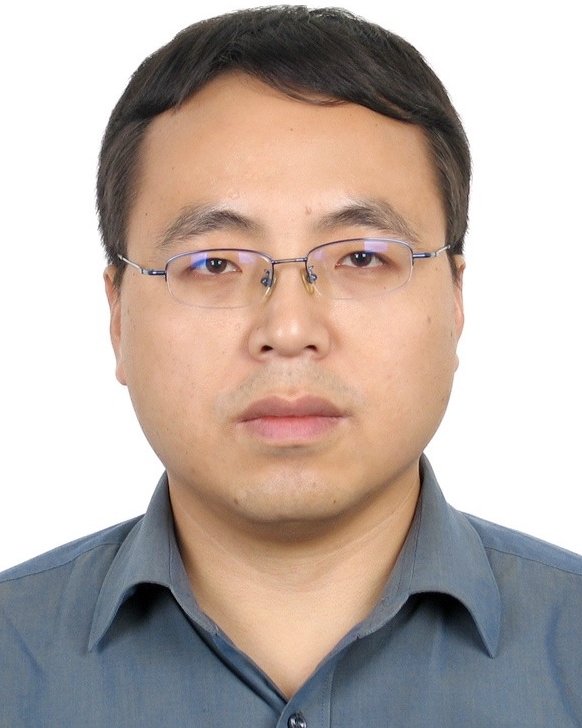}}]
{\newline \newline Junchao Zhang} is a software engineer at Argonne National Laboratory.
 He received his Ph.D. in computer science from the Chinese Academy of Sciences, Beijing, China.
 He is a PETSc developer and  works mainly on communication and GPU support in PETSc.
\end{IEEEbiography}

\vspace{-43pt}
\begin{IEEEbiography}[{\includegraphics[width=1in,height=1.25in,clip,keepaspectratio]{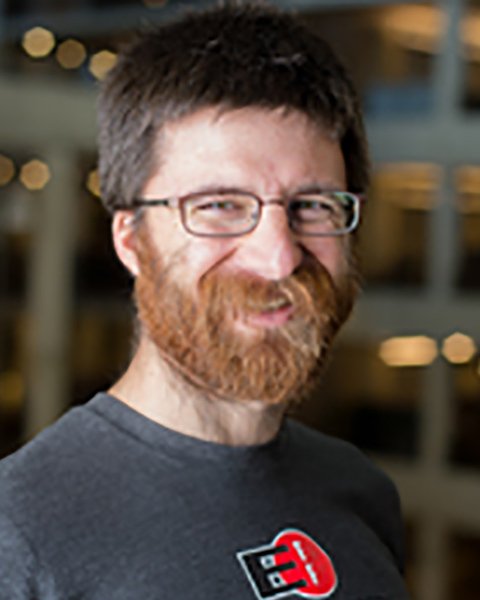}}]
{\newline Jed Brown} is an assistant professor of computer science at the University of Colorado Boulder. He received his Dr.Sc. from ETH Z\"urich and BS+MS from the University of Alaska Fairbanks. He is a maintainer of PETSc and leads a research group on fast algorithms and community software for physical prediction, inference, and design.
\end{IEEEbiography}

\vspace{-43pt}
\begin{IEEEbiography}[{\includegraphics[width=1in,height=1.25in,clip,keepaspectratio]{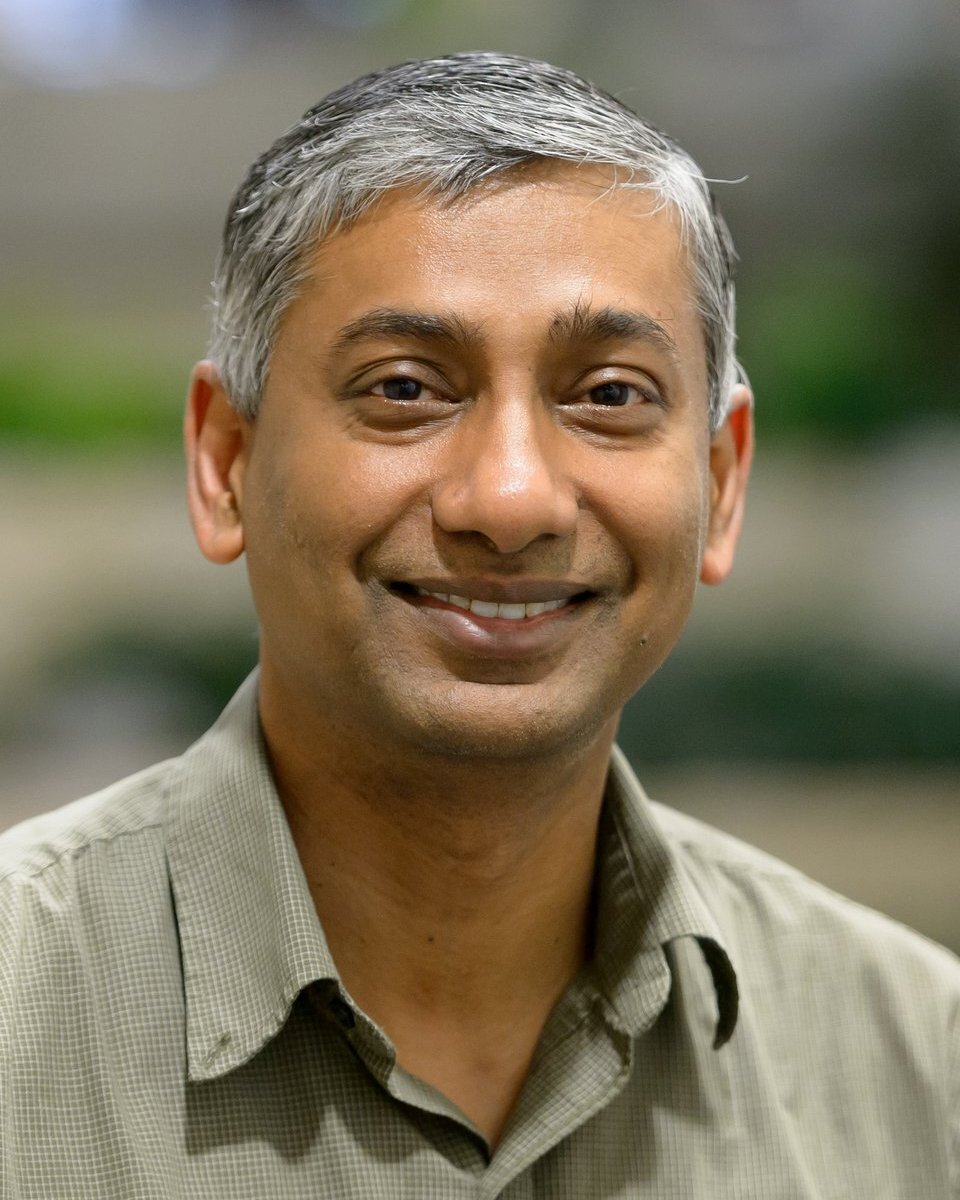}}]
{\newline Satish Balay} is a software engineer at Argonne National Laboratory. He received his M.S. in computer science from Old Dominion University. He is a developer of PETSc.
\end{IEEEbiography}

\vspace{-43pt}
\begin{IEEEbiography}[{\includegraphics[width=1in,height=1.25in,clip,keepaspectratio]{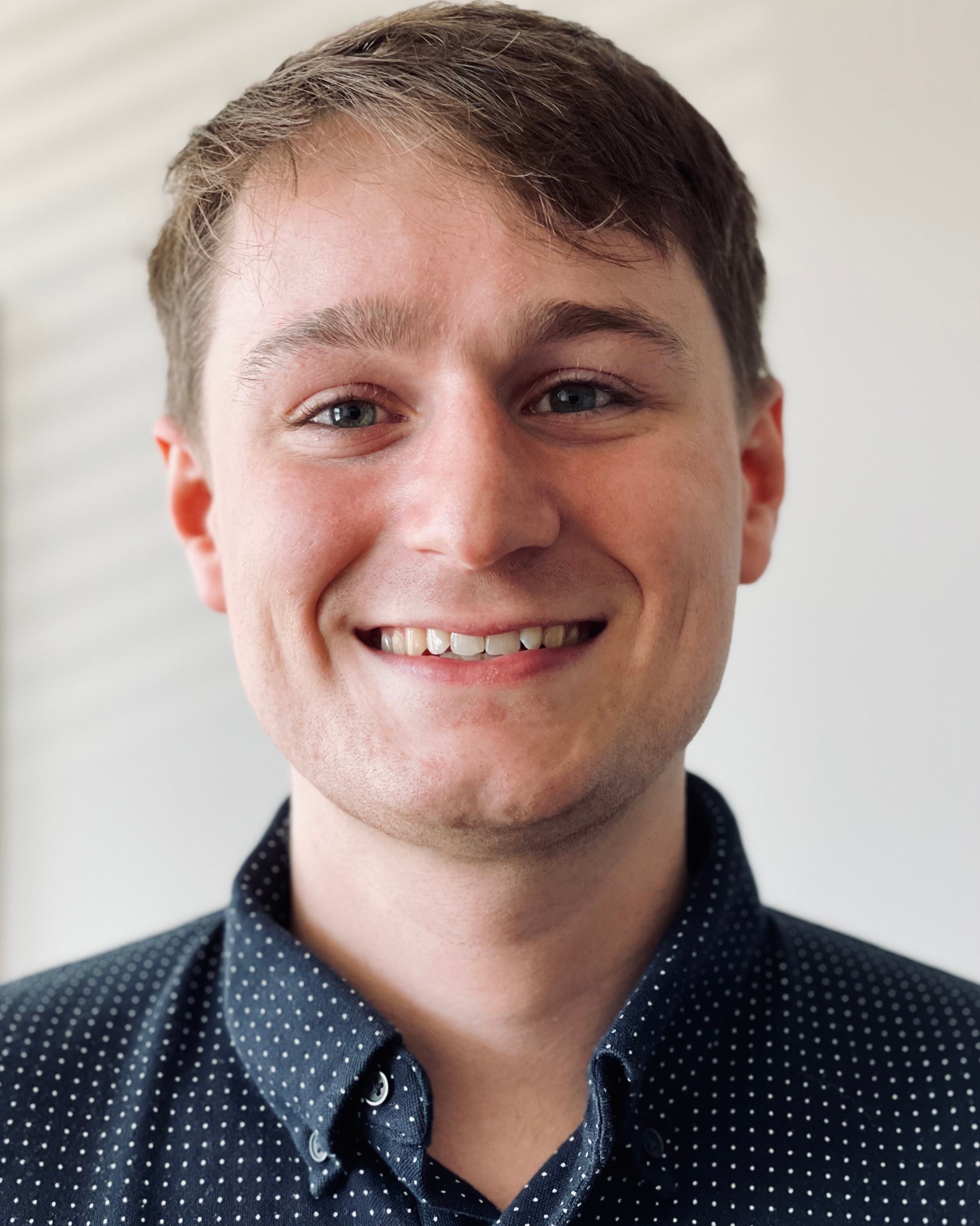}}]
{\newline Jacob Faibussowitsch} is a Ph.D. student in mechanical engineering and computational science and engineering at the University of Illinois at Urbana-Champaign, where he also received his B.S. His work focuses on high-performance scalable fracture mechanics at the Center for Exascale-Enabled Scramjet Design. He is a developer of PETSc.
\end{IEEEbiography}

\vspace{-43pt}
\begin{IEEEbiography}[{\includegraphics[width=1in,height=1.25in,clip,keepaspectratio]{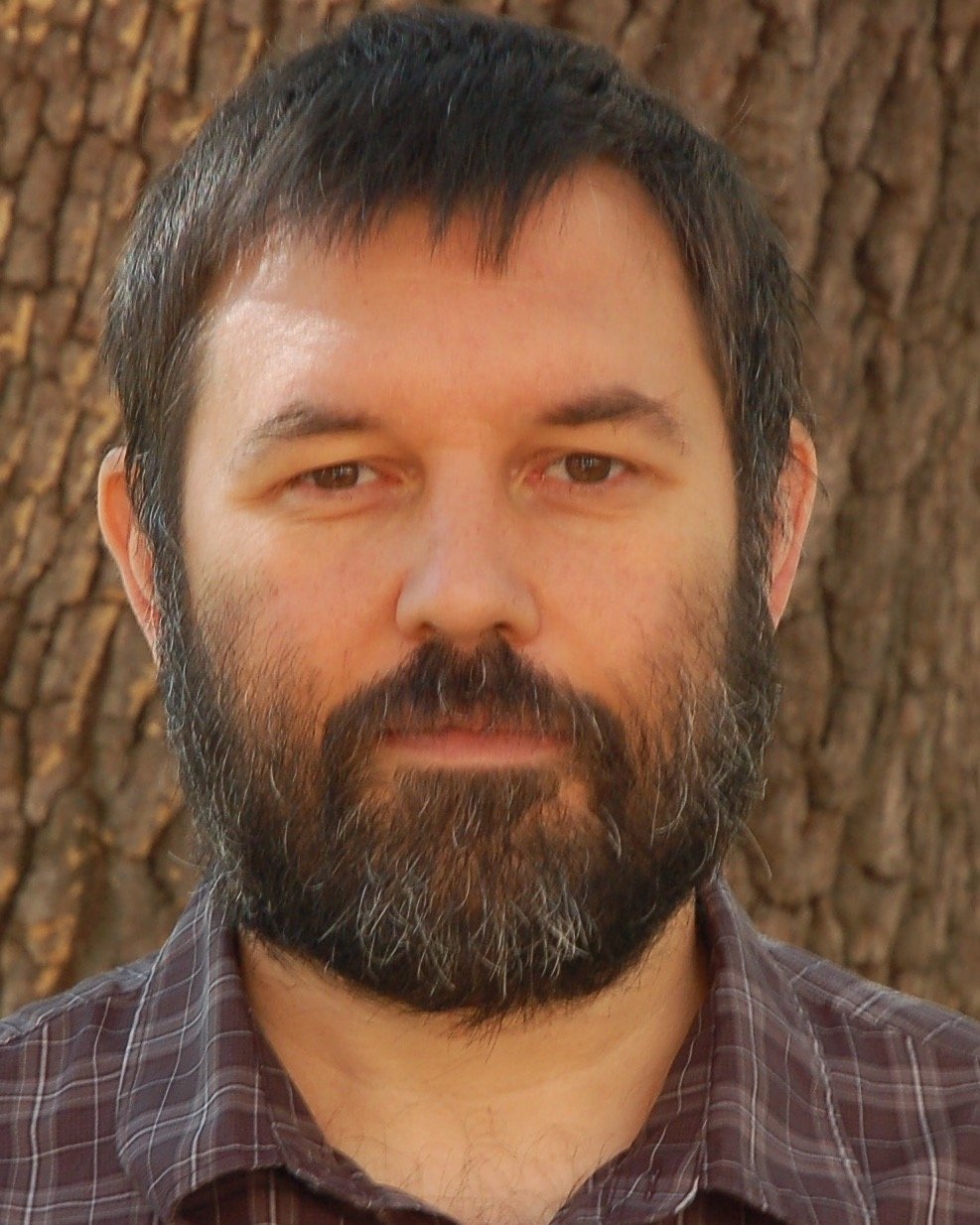}}]
{\newline Matthew Knepley} is an associate professor in the University at Buffalo. He received his Ph.D. in computer science from Purdue University and his B.S. from Case Western Reserve University. His work focuses on computational science, particularly in geodynamics, subsurface flow, and molecular mechanics. He is a maintainer of PETSc.
\end{IEEEbiography}

\vspace{-43pt}
\begin{IEEEbiography}[{\includegraphics[width=1.0\textwidth,clip,keepaspectratio]{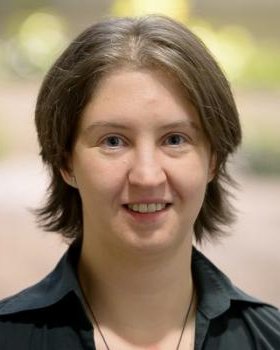}}]
{\newline Oana Marin} is an assistant applied mathematics specialist at Argonne National Laboratory. She received her Ph.D. in theoretical numerical analysis at the Royal Institute of Technology, Sweden. She is an applications-oriented applied mathematician who works on numerical discretizations in computational fluid dynamics, mathematical modeling, and data processing.
\end{IEEEbiography}

\vspace{-43pt}
\begin{IEEEbiography}[{\includegraphics[width=1in,height=1.25in,clip,keepaspectratio]{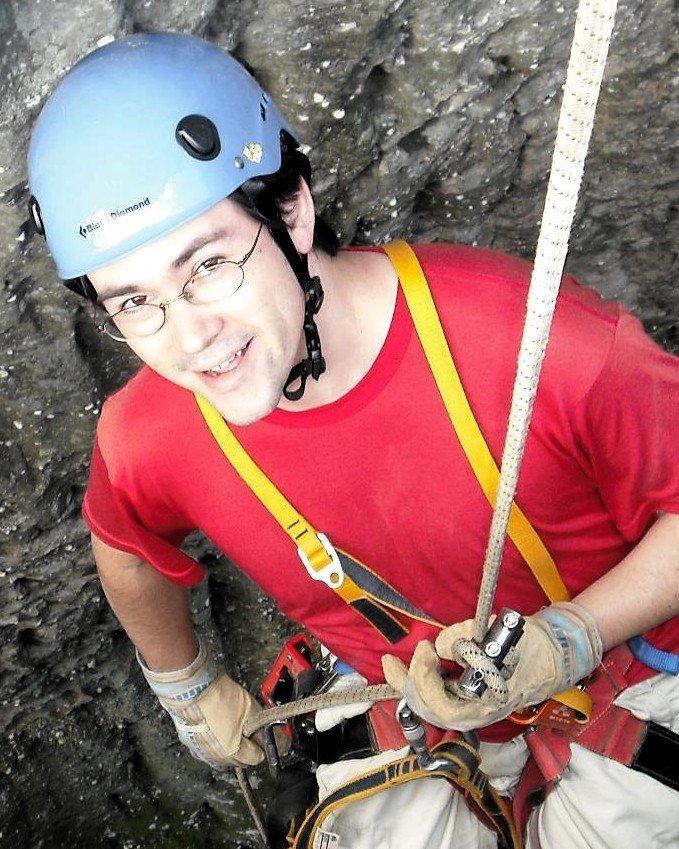}}]
{\newline Richard Tran Mills} is a computational scientist at Argonne National Laboratory. His research spans high-performance scientific computing, machine learning, and the geosciences.
He is a developer of PETSc and the hydrology code PFLOTRAN.
He earned his Ph.D. in computer science at the College of William and Mary, supported by a U.S. Department of Energy Computational Science Graduate Fellowship.
\end{IEEEbiography}

\vspace{-43pt}
\begin{IEEEbiography}[{\includegraphics[trim=25 120 40 15, width=1in,height=1.25in,clip,keepaspectratio]{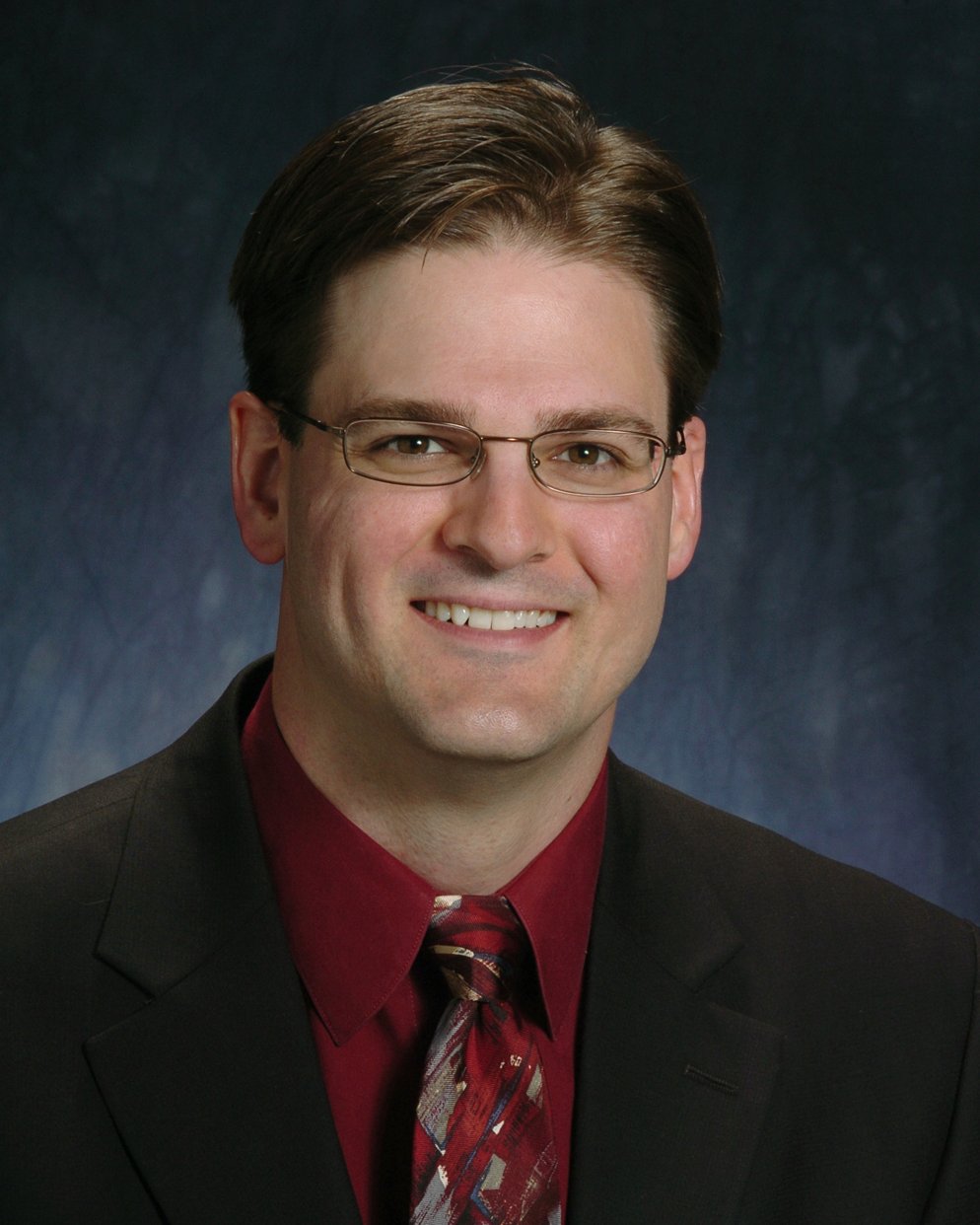}}]
{\newline Todd Munson} is a senior computational scientist
at Argonne National Laboratory
and the Software Ecosystem and Delivery Control
Account Manager for the U.S.\ DOE Exascale Computing Project.
His interests range from numerical 
methods for nonlinear optimization and variational 
inequalities to workflow optimization for online 
data analysis and reduction.
He is a developer of the Toolkit for Advanced Optimization.
\end{IEEEbiography}

\vspace{-46pt}
\begin{IEEEbiography}[{\includegraphics[width=1in,height=1.25in,clip,keepaspectratio]{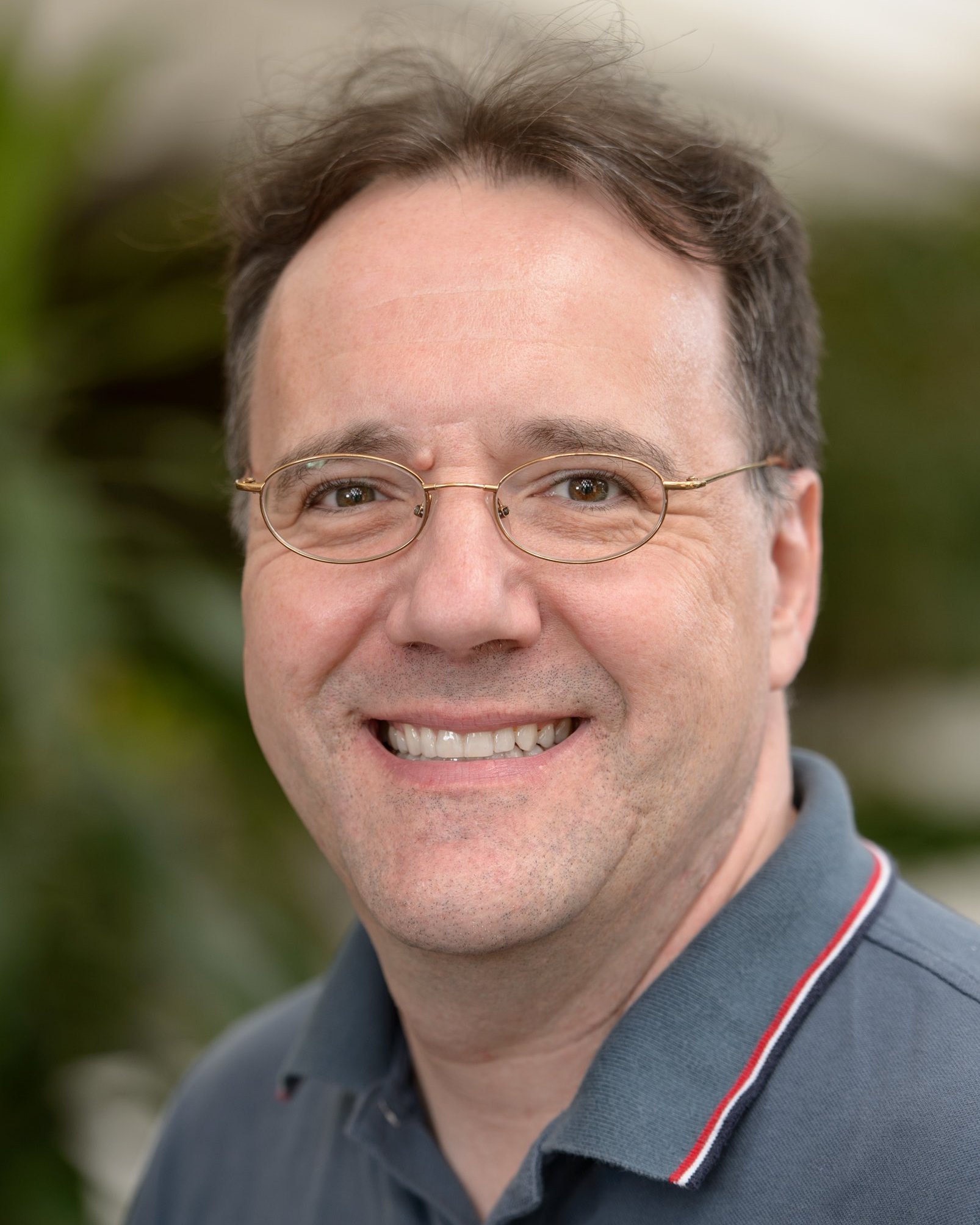}}]
{\newline Barry F. Smith} is an Argonne National Laboratory Associate. 
He is one of the original developers of the PETSc numerical solvers library.
He earned his Ph.D. in mathematics at the Courant Institute.
\end{IEEEbiography}

\vspace{-46pt}
\begin{IEEEbiography}[{\includegraphics[width=1in,height=1.25in,clip,keepaspectratio]{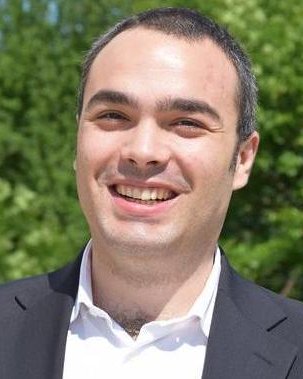}}]
{\newline Stefano Zampini} is a research scientist in the Extreme Computing Research Center of
King Abdullah University for Science and Technology (KAUST), Saudi Arabia. He received his Ph.D. in applied mathematics from the University of Milano Statale.
He is a developer of PETSc.
\end{IEEEbiography}

\onecolumn
\centering
\framebox{
\parbox{4in}{
The submitted manuscript has been created by UChicago Argonne, LLC, Operator of Argonne
National Laboratory (``Argonne''). Argonne, a US Department of Energy Office of Science
laboratory, is operated under Contract No. DE-AC02-06CH11357. The US Government retains
for itself, and others acting on its behalf, a paid-up nonexclusive, irrevocable worldwide
license in said article to reproduce, prepare derivative works, distribute copies to the
public, and perform publicly and display publicly, by or on behalf of the Government.
The Department of Energy will provide public access to these results of federally
sponsored research in accordance with the DOE Public Access
Plan. \url{http://energy.gov/downloads/doe-public-accessplan}
}}

\end{document}